\documentclass[aps,prd,twocolumn,groupedaddress,showpacs,amsmath,amssymb]{revtex4}
\usepackage{epsfig,graphicx}
\usepackage{amsmath,amssymb}
\usepackage{amsfonts}

 \newcommand{\be}[1]{\begin{equation}\label{#1}}
    \newcommand{\ba}[1]{\begin{eqnarray}\label{#1}}

    \newcommand{\pa}[1]{\left(#1\right)}
    \newcommand{\paq}[1]{\left[#1\right]}
    
    \def\be{\begin{equation}}
    \def\ba{\begin{eqnarray}}
    \def\ee{\end{equation}}
    \def\ea{\end{eqnarray}}

\begin{document}

\title{Reconstruction of Scalar Potentials in Modified Gravity Models}

\author{A.Yu. Kamenshchik$^{1,2,3}$, A. Tronconi$^{1,2}$, G. Venturi$^{1,2}$ and S.Yu.~Vernov$^{4,5}$}
\affiliation{$^{1}$ Dipartimento di Fisica e Astronomia, Universit\`a degli Studi di
Bologna, Via Irnerio 46, I--40126 Bologna -- Italy}
\affiliation{$^{2}$ INFN, Sezione di Bologna, Via Irnerio 46, I--40126 Bologna -- Italy}
\affiliation{$^{3}$ L.D. Landau Institute for Theoretical Physics of the Russian
Academy of Sciences, Kosygin str. 2, 119334 Moscow -- Russia}
\affiliation{$^{4}$ Skobeltsyn Institute of Nuclear Physics, Lomonosov  Moscow State University, Leninskie Gory 1, 119991, Moscow -- Russia}
\affiliation{$^{5}$ Instituto de Ciencias del Espacio (ICE/CSIC) and  Institut d'Estudis Espacials de Catalunya, Campus UAB, Fac. Ci\`encies, Torre C5-Parell-2a pl., E-08193, Bellaterra, Barcelona -- Spain}

\begin{abstract}
We employ the superpotential technique for the reconstruction of cosmological models with a non-minimally coupled scalar field evolving on a spatially flat Friedmann--Robertson--Walker background. The key point in this method is that the Hubble parameter is considered as a function of the scalar field and this allows one to reconstruct the scalar field potential and determine the dynamics of the field itself, without \textit{a priori} fixing the Hubble parameter as a function of time or of the scale factor. The scalar field potentials that lead to de Sitter or asymptotic de Sitter solutions, and those that reproduce the cosmological evolution given by Einstein-Hilbert action plus a barotropic perfect fluid, have been obtained.
\end{abstract}
\maketitle
\section{Introduction}
There are two possible ways of describing the accelerating expansion of the Universe \cite{cosmic}.
One can assume that General Relativity is the
correct theory of gravity and introduce a smoothly distributed, slowly varying cosmic fluid with negative pressure called dark energy~\cite{DE_rev}.  Alternatively, one can modify the theory of gravity~\cite{Fujii_Maeda,NO-rev,Book-Capozziello-Faraoni,CL}. Some of these modified gravity models, for example, $F(R)$ gravity models, can then be mapped into general relativity with additional scalar fields by a suitable conformal transformation of the metric (see, e.g.,~\cite{Book-Capozziello-Faraoni,CL,Felice_Tsujikawa}).
It is well known that scalar fields play an essential role in modern cosmology since they are possible candidates for the role of the inflaton field driving inflation in the early universe \cite{inflation,inflation2} and of the dark energy substance \cite{DE_rev}. The Universe expansion history, unifying early-time inflation and late-time acceleration, can be realized in scalar-tensor gravity~\cite{ENOSF}.

The modified gravity models, as well as cosmological models with scalar fields, often include some function that
cannot be deduced from the fundamental theory. It is then a natural question to ask, why some specific form of such a function is chosen and what is the physical motivation for this choice. The technique of reconstruction has attracted the attention of researchers for a long time and has been developed for different cosmological models~\cite{Starobinsky:1998fr}--\cite{LocalScalarTensor}. For example, the reconstruction of potentials for models
with minimally coupled scalar fields has been studied in~\cite{Starobinsky:1998fr}--\cite{NoOdin2007}
and, in the
two-field case,  in~\cite{AKV2,Andrianov:2007ua,ABV}, while a similar procedure for tachyon models was discussed in~\cite{Padmanabhan:2002cp,Feinstein:2002aj,Gorini:2003wa,Rotova}. The reconstruction of potentials for scalar fields non-minimally coupled to gravity was considered in \cite{Boisseau:2000pr,KTV11042125,Qui}.
We can also mention the reconstruction procedures in models with non-minimally coupled Yang--Mills fields~\cite{BNO}, in $F(R)$ and Gauss--Bonnet gravity models~\cite{NoOdin2007,delaCruzDombriz:2006fj,FR_reconsruction}, in $F(T)$ models, with $T$ being the torsion scalar~\cite{FT_Reconstruction}, in the nonlocal gravity model~\cite{DW} and its local formulation~\cite{LocalScalarTensor}.

Gravity models with non-minimally coupled scalar fields are actively studied in cosmology~\cite{Cooper:1982du,induced,Cerioni:2009kn,Cerioni:2010ke,Tronconi:2010pq,Kamenshchik:2012rs,Sami:2012uh,CervantesCota:2010cb} (see also~\cite{Book-Capozziello-Faraoni} and references therein). In particular, the models with the Hilbert--Einstein term plus the term proportional to the Ricci scalar multiplied by the square of the scalar field were intensively studied in inflationary \cite{nonmin-inf} and in quantum \cite{nonmin-quant} cosmology.

Generally these models are described by the following action:
\begin{equation}
\label{action}
S=\int d^4 x \sqrt{-g}\left[ U(\sigma)R-\frac12g^{\mu\nu}\sigma_{,\mu}\sigma_{,\nu}+V(\sigma)\right],
\end{equation}
where $U(\sigma)$ and $V(\sigma)$ are differentiable functions of the scalar field $\sigma$. We use the
signature $(+,-,-,-)$, and $g$ is the determinant of the metric tensor
$g_{\mu\nu}$.

The reconstruction procedure for the induced gravity models ($U(\sigma)=\xi\sigma^2$ where $\xi$ is a constant) has been proposed in~\cite{KTV11042125}.
In such a case it has been shown that one can linearize all the differential equations that should be solved in the reconstruction procedure to get the potential corresponding to a given cosmological evolution. This property allows one to obtain the explicit forms of potentials reproducing the dynamics of a flat Friedmann--Robertson--Walker (FRW) universe, driven by barotropic perfect fluids, by a Chaplygin gas \cite{Chap} and by a modified Chaplygin gas~\cite{KTV11042125} (The term modified Chaplygin gas was proposed in \cite{mod-Chap} while similar equations of state were considered in \cite{Barrow} and
\cite{we-Chap1}).

In this paper, we consider another reconstruction procedure for the models described by the above action~(\ref{action}).
Such a method is similar to the Hamilton--Jacobi method (also known as the superpotential method) and is commonly applied to cosmological models with scalar fields~\cite{Superpotential,AKV2,Andrianov:2007ua,ABV,Rotova} as well as in brane models~\cite{Extradim}.
The key point in this method is that the Hubble parameter is considered as a function of the scalar field $\sigma$.

The two methods described above supplement each other and together allow one to construct different cosmological models with some required properties.
In particular, the reconstruction procedure illustrated in~\cite{KTV11042125} is useful in order to obtain the potential and the explicit evolution of the scalar field when the Hubble parameter is given explicitly as a function of the cosmic time $t$ or of the scale factor $a$. The superpotential method also allows one to do so (see Section 4), but for $U(\sigma)=\xi\sigma^2$ the procedure proposed in~\cite{KTV11042125} is simpler. At the same time, however, the superpotential variant of the reconstruction procedure allows one to construct models and their exact solutions without any knowledge of the exact behavior of the Hubble parameter. For example, in Section 6 we describe the dynamics of a model with the Hubble parameter evolving to a nonzero constant value at late times. Such a method also allows one to calculate the function $U(\sigma)$, when the Hubble parameter is given as function of $\sigma$ and the behavior of $\sigma$ as a function of the cosmic time $t$ (or the
  scale factor $a$) is known. Specifically, we assume that $\frac{d\sigma}{dt}=F(\sigma)$, where the function $F$ is given. So, the superpotential procedure allows one to reconstruct not only the potential, but also either $U(\sigma)$ or the Hubble parameter, if the evolution of $\sigma$ is given. Furthermore we can use this method to reconstruct scalar field potentials with some given property. For example, in Section~7 we find the conditions on the functions $U(\sigma)$ and $F(\sigma)$ for which the potential obtained is polynomial.

By using this method we find new exact solutions in the induced gravity case.
We also applied the procedure to cosmological models with a non-minimally coupled scalar field described by (\ref{action}) and $U(\sigma)=\xi\sigma^2+J$, where $J$ is a constant. In this physically
important case, the method, proposed in~\cite{KTV11042125}, requires one to find solutions of second order nonlinear differential equations, whereas on using the superpotential method one only has to solve linear differential equations and first order autonomous nonlinear differential equations. Thus, the solutions can always be obtained at least in quadratures. One important case treated in detail is the reconstruction of de Sitter solutions. In such a case, the method proposed in~\cite{KTV11042125} is still useful for arbitrary $U(\sigma)$ (see Section 3).

The paper is organized as follows.
In Section~2, we review the basic equations for a gravity model with a non-minimally coupled scalar field in a FRW background and we describe the algorithm of the superpotential reconstruction procedure.
In Section~3, we study the de Sitter solutions by using the
methods proposed in~\cite{KTV11042125} and in this paper and compare results.
In Section~4, we find the non-minimally coupled models reproducing the dynamics of a flat FRW universe, driven by barotropic perfect fluids.
In Section~5, we assume that the scalar field is proportional to the hyperbolic tangent (or cotangent) of the cosmic time and consider possible behaviors of the Hubble parameter and the corresponding potentials.
In Section~6, we consider the induced gravity model and find solutions and potentials with a Hubble parameter that evolves toward a constant value at late times.
In Section~7, we find conditions on the functions $U(\sigma)$ and $F(\sigma)$ sufficient to obtain a model with a polynomial potential.
Finally Section~8 is devoted to the conclusions.

\section{Cosmological models with non-minimally coupled scalar field}
\subsection{Homogeneous equations}
Consider the evolution of a homogeneous scalar field on a spatially
flat FRW universe with the metric
\begin{equation*}
ds^2=dt^2-a^2(t)\left(dx_1^2+dx_2^2+dx_3^2\right).
\end{equation*}
The Einstein equations derived from the variation of the action (\ref{action}) have the following form\footnote{In~\cite{KTV11042125}, the authors have used a convenient choice of Newton's constant $G$, $8\pi G/3 = 1$, and dimensionless quantities. Alternatively, one can assume that $G$ is a part of $U(\sigma)$ and do not fix the value of this constant.}~\cite{KTV11042125}:
\begin{equation}
\label{equ00}
6UH^2+6\dot U H=\frac{1}{2}\dot\sigma^2+V,
\end{equation}
\begin{equation}
\label{equ11}
2U\left(2\dot H+3H^2\right)+4\dot U H+2\ddot U +\frac{1}{2}\dot\sigma^2-V=0,
\end{equation}
where differentiation with respect to time $t$ is denoted by a dot, and the Hubble parameter is  the logarithmic derivative of the scale factor:
$H=\dot a/a$.
The variation of action (\ref{action}) with respect to $\sigma$ gives
the Klein--Gordon equation
\begin{equation}
\ddot \sigma+3H\dot\sigma+V_{,\sigma}=6\left(\dot H +2H^2\right)U_{,\sigma}\,,
\end{equation}
where the subscript $ _{,\sigma}$ indicates the derivative with respect to the scalar field
$\sigma$.
 Combining  Eqs. (\ref{equ00}) and (\ref{equ11}) we obtain
\begin{equation}
\label{equ01}
4U\dot H-2\dot U H+2\ddot U +\dot\sigma^2=0.
\end{equation}
This equation plays a key role in the reconstruction procedure.
\subsection{The Hubble parameter as a function (superpotential) of the scalar field.}
Let $H=Y(\sigma)$, then (\ref{equ01}) takes the following form:
\begin{equation}
\label{equ01W}
4UY_{,\sigma}\dot\sigma-2YU_{,\sigma}\dot\sigma+2U_{,\sigma\sigma}\dot\sigma^2+2U_{,\sigma}\ddot\sigma+\dot\sigma^2=0.
\end{equation}
Let us  introduce the function $F(\sigma)$ defined as
\begin{equation}
\label{equsigma}
\dot \sigma=F(\sigma).
\end{equation}
Substituting $\dot\sigma$ and $\ddot \sigma=F_{,\sigma}F$ into (\ref{equ01W}), we obtain
\begin{equation}
\label{equa}
4UY_{,\sigma}+2(F_{,\sigma}-Y)U_{,\sigma}+(2U_{,\sigma\sigma}+1)F=0.
\end{equation}
Equation (\ref{equa}) contains three functions. If two of them are given,
then the third one can be found as the solution of a linear differential
equation.  Let us note that Eq.~(\ref{equa}) is a first order
differential equation for both  $Y$ and $F$.

If, for example, $U$ and $F$ are given, then Eq.~(\ref{equa}) can easily be  integrated to obtain $Y(\sigma)$:
\begin{equation}
\label{Ysigma}
Y(\sigma)=-\left[\int\limits^{\sigma}\frac{2F_{,\tilde\sigma}U_{,\tilde\sigma}+(2U_{,\tilde\sigma\tilde\sigma}+1)F}{4U^{3/2}}
\,d\tilde \sigma+c_0\right]\!\sqrt{U(\sigma)},
\ee
where  $c_0$ is an integration constant.
In a similar fashion, for given $Y(\sigma)$ and $U(\sigma)$, we obtain $F(\sigma)$ by integrating~(\ref{equa}):
\begin{equation}
\label{SolF}
    F(\sigma)=\left[ \int\limits^\sigma \frac{U_{,\tilde{\sigma}}Y-2U Y_{,\tilde{\sigma}}}{U_{,\tilde{\sigma}}}  e^{\Upsilon} d\tilde{\sigma}+\tilde c_0\phantom{\frac{A}{B}}\!\!\!\!\!
    \right]e^{-\Upsilon(\sigma)},
\end{equation}

where
\begin{equation*}
    \Upsilon(\sigma)\equiv\frac{1}{2}\int\limits^{\sigma}\frac{2U_{,\tilde\sigma\tilde\sigma}+1}
    {U_{,\tilde\sigma}}\,d\tilde\sigma
\end{equation*}
and $\tilde c_{0}$ is an integration constant.

The potential $V(\sigma)$ can then be obtained from (\ref{equ00}):
\begin{equation}
\label{potentialV}
V(\sigma)=6UY^2+6U_{,\sigma}FY-\frac{1}{2}F^2.
\end{equation}
To find the time evolution $\sigma(t)$ we finally integrate Eq.~(\ref{equsigma}), which can always be solved in quadratures.

The evolution of the Hubble parameter $H$ and of the scalar field $\sigma$ can be expressed through either  the cosmic time $t$ or the scale
factor $a$. In the latter case it is more convenient to adopt the dimensionless parameter $N\equiv \ln
(a/a_{0})$ as an independent variable. In terms of $N$ one has to reexpress the time derivatives as
\begin{equation}
\label{ttoN} \frac{d}{dt}=H\frac{d}{dN}\quad {\rm and}\quad
\frac{d^{2}}{dt^{2}}=H^{2}\frac{d^{2}}{dN^{2}}+\frac{1}{2}\left(H^{2}\right)'\frac{d}{dN}\,,
\end{equation}
where the prime denotes the derivative with respect to~$N$. Equation~(\ref{equa}) then takes the following form:
\begin{equation}
\label{startN}
2U(H^{2})'-2U'H^{2}+2U''H^{2}+(H^{2})'U'+(\sigma')^{2}H^{2}=0.
\end{equation}
On defining the functions $W(\sigma)$ and $\tilde{F}(\sigma)$ as
follows
\begin{equation} \label{WN} W(\sigma)\equiv H^{2} \end{equation}
and
\begin{equation} \label{FN}
\tilde{F}(\sigma)\equiv \sigma',
\end{equation}
we find
\begin{equation} \label{dH2}
(H^{2})'=\frac{dH^{2}}{d\sigma}\frac{d\sigma}{dN}\equiv W_{,\sigma} \tilde{F},
\end{equation}
 \begin{equation} \label{dU} U'=U_{,\sigma}\tilde{F},\qquad
U''=\pa{U_{,\sigma\sigma}\tilde{F}+U_{,\sigma}\tilde{F}_{,\sigma}}\tilde{F}
\end{equation}
and Eq. (\ref{startN}) can then be cast in the form:
\begin{eqnarray} \label{starts}
&& 2UW_{,\sigma}-2U_{,\sigma}W+2\pa{U_{,\sigma\sigma}\tilde{F}+U_{,\sigma}\tilde{F}_{,\sigma}}W\nonumber\\
&&+W_{,\sigma}U_{,\sigma}\tilde{F}+\tilde{F}W=0.
\end{eqnarray}
If, for example, $\tilde{F}(\sigma)$ and $U(\sigma)$ are given, the function $W(\sigma)$
can be determined by integrating
\begin{equation} \label{solW}
\frac{W_{,\sigma}}{W}=\frac{\paq{2U_{,\sigma}-2\pa{U_{,\sigma\sigma}\tilde{F}+U_{,\sigma}\tilde{F}_{,\sigma}}-\tilde{F}}}
{\pa{2U+U_{,\sigma}\tilde{F}}}
\end{equation}
and finally the potential can be reconstructed by using
\begin{equation} \label{recpN}
V=W\paq{6U+6U_{,\sigma}\tilde F-\frac{1}{2}\tilde F^{2}}.
\ee
Equation (\ref{solW}) connects the functions $W$, $U$, and $\tilde{F}$ and is equivalent to Eq.~(\ref{equa}).
Indeed, on substituting $F=Y\tilde{F}$ into Eq.~(\ref{equa}) we get
\begin{equation}
2\frac{Y_{,\sigma}}{Y}=\frac{\paq{2U_{,\sigma}-2\pa{U_{,\sigma\sigma}\tilde{F}+U_{,\sigma}\tilde{F}_{,\sigma}}-\tilde{F}}}
{\pa{2U+U_{,\sigma}\tilde{F}}}.
\end{equation}
The latter equation is equivalent to Eq.~(\ref{solW}), because $W(\sigma)=Y^2(\sigma)$.
We thus observe that the superpotential method allows one to unify the reconstruction procedures for the functions in (\ref{equa})
given as functions of $t$ or of $N$. Indeed, one can start from Eq.~(\ref{equa}) and find suitable functions $Y$, $F$ and $U$ satisfying it and, after that, it is trivial to get $W$ and~$\tilde{F}$.

\subsection{Quadratic non-minimally coupled models}
In this paper, we consider
\begin{equation}
\label{U2poly}
U(\sigma)=\xi\sigma^2+C_1\sigma+J_0.
\end{equation}
The case of induced gravity, for which the reconstruction procedure has been proposed in~\cite{KTV11042125}, corresponds to a particular choice of $U(\sigma)$ given by $C_1=0$ and $J_0=0$.

Note that the quadratic polynomial (\ref{U2poly}) can be rewritten as
\begin{eqnarray}
\label{U2poly1}
&U(\sigma)=\xi(\sigma-\sigma_0)^2+J, &\nonumber\\
&\mbox{with }\;\sigma_0={}-\frac{C_1}{2\xi}, \; J=J_0-\frac{C_1^2}{4\xi}.&
\end{eqnarray}
In this paper, we consider the case $\sigma_0=0$ (equivalent to $C_1=0$). Let us observe, however, that such a restriction does not lead to any loss of generality because any solution found with $\sigma_{0}=0$ can be extended to the $\sigma_{0}\neq0$ case simply through the change of variable $\sigma\rightarrow \sigma-\sigma_{0}$.

For $U(\sigma)$, given by (\ref{U2poly}) with $\sigma_{0}=0$, we obtain from (\ref{Ysigma}) and (\ref{SolF}):
\begin{eqnarray}
\label{equasolY}
Y(\sigma)&=&{}-\left[\int\limits^{\sigma}\frac{4\xi\tilde\sigma F_{,\tilde{\sigma}}+(4\xi+1)F}{4(\xi\tilde\sigma^2+J)^{3/2}}\,d\tilde\sigma+ c_0\right]\nonumber\\
&\times&\sqrt{\xi\sigma^2+J}\,,
\end{eqnarray}
\begin{eqnarray}
\label{equasolF}
F(\sigma)&=&\left\{\int\limits^{\sigma}\left[\tilde\sigma Y-\left(\tilde{\sigma}^2+\frac J\xi\right)Y_{,\tilde\sigma}\right]\tilde\sigma^{\frac{1}{4\xi}}\,d\tilde\sigma+\tilde c_0\right\}\nonumber\\
&\times& \sigma^{-\frac{1+4\xi}{4\xi}}\,.
\end{eqnarray}

Let us note that the behavior of the Hubble parameter can be very different in a model with zero and nonzero $J$. To illustrate this point let us consider the example with an arbitrary linear function $F(\sigma)$:
\begin{equation}
F=f_1\sigma+f_0,
\end{equation}
where $f_i$ are constants.
From Eq.~(\ref{equasolY}), we get
\begin{eqnarray} \label{spl1}
&Y_J(\sigma)=\frac{(8\xi+1)f_1}{4\xi}-\frac{(4\xi+1) f_0}{4J}\sigma+c_0\sqrt{\xi\sigma^2+J}\,,&\\
&\mbox{with}\quad J\neq 0,&\nonumber
\ea
and
\begin{eqnarray} \label{spl2}
&Y_0(\sigma)=\frac{f_1}{4\xi}+2f_1+\frac{(4\xi+1) f_0}{8\xi\sigma}+c_0\sigma\,,&\\
&\mbox{with}\quad J=0.&\nonumber
\ea
Therefore, if $f_0\neq 0$ and $\xi\neq -1/4$, then $Y_0(\sigma)$ is not a limit of $Y_J(\sigma)$ at $J\rightarrow 0$.

\section{Models with de Sitter solutions}
In this section we shall consider the general form of the potential $V(\sigma)$, which leads to the existence of the
de Sitter solution
\begin{equation}
H = H_0 = \mathrm{const}
\label{const}
\end{equation}
for an arbitrary non-minimal coupling $U(\sigma)$. In this case, it is convenient to apply the technique
developed in~\cite{KTV11042125}.

First of all let us notice that if the scalar field $\sigma$ does not depend on time $t$ (or, in other words, on the
scale factor $a$), then all the considerations presented in the end of the second section of paper \cite{KTV11042125}
are still valid and we obtain immediately the de Sitter solution for the model with the potential (see Eq. (31) of the above cited paper)
\begin{equation}
V(\sigma) = V_0 U^2(\sigma),
\label{constant1}
\end{equation}
where $V_{0}$ is an arbitrary positive constant. The constant value of the Hubble parameter is then (see Eq. (\ref{equ00}))
\begin{equation}
H_0 = \pm\sqrt{\frac{V}{6U}} =\pm \sqrt{\frac{V_0 U}{6}}.
\label{constant2}
\end{equation}

For the case of a time-dependent scalar field $\sigma$, we use Eq.~(19) from paper \cite{KTV11042125}, and we rewrite it in terms of $N\equiv \ln (a/a_{0})$
\begin{equation}
\sigma''-\sigma' + \sigma' \frac{H'}{H} + (\sigma')^2\left(\frac{1+2U_{,\sigma\sigma}}{2U_{,\sigma}}\right) + \frac{2UH'}{U_{_,\sigma}H} = 0.
\label{old}
\end{equation}
For the de Sitter case, this equation reduces to
\begin{equation}
\sigma''-\sigma'= {}-\left(\frac{1+2U_{,\sigma\sigma}}{2U_{,\sigma}}\right)(\sigma')^2.
\label{old1}
\end{equation}
Using $\frac{d}{dN}=\sigma' \frac{d}{d\sigma}$, after some algebra, one can integrate both sides of Eq. (\ref{old1}) and obtain
\begin{eqnarray} \label{old23}
\ln \sigma'-N&=&{}-\int\limits^{\sigma} \pa{\frac{1+2U_{,\tilde\sigma\tilde\sigma}}{2U_{,\tilde\sigma}}}d\tilde{\sigma}+\tilde{B_{0}}\nonumber\\
&=&{}-\int\limits^{\sigma} \frac{d\tilde \sigma}{2U_{,\tilde\sigma}}-\ln\left(U_{,\sigma}\right)+\tilde B_{0}.
\ea
where $\tilde B_{0}$ is an integration constant.

We now exponentiate and integrate the above equation to find  $\exp(N)$ as a function of the scalar field $\sigma$:
\begin{equation}
\exp(N) = \frac{1}{B_0}\int\limits^{\sigma} U_{,\sigma_{*}}\exp\left(\frac12\int\limits^{\sigma_{*}}  \frac{d\tilde \sigma}{U_{,\tilde\sigma}}\right)d\sigma_{*}  + B_1,
\label{old7}
\end{equation}
where $B_{0}=\exp (\tilde B_{0})$ and $B_1$ is an arbitrary constant.

Consequently we obtain $\sigma'$ as a function of $\sigma$:
\begin{eqnarray}
\label{G}
\!\!\!\!\sigma' &=& \frac{\exp\left(-\frac12\int\limits^{\sigma} \frac{d\tilde \sigma }{U_{,\tilde\sigma}}\right)}{U_{,\sigma}}\nonumber\\
&\times&\left[B_1B_0+
\int\limits^{\sigma} U_{,\sigma_{*}} \exp \left(\frac12\int\limits^{\sigma_{*}} \frac{d\tilde\sigma }{U_{,\tilde\sigma}}\right) d\sigma_{*} \right]
\end{eqnarray}
and the corresponding potential can be found by substituting (\ref{G}) and (\ref{const}) into Eq.~(\ref{recpN}).

For the particular case
\begin{equation}
U(\sigma) =\xi \sigma^2+ J,
\label{particular}
\end{equation}
we obtain
\begin{equation}
\sigma' = \frac{4\xi}{1+8\xi}\sigma+\frac{B_1B_0}{2\xi} \sigma^{-\frac{1+4\xi}{4\xi}}.
\label{B1}
\end{equation}
Substituting (\ref{B1}) into (\ref{equ00}) we get
\begin{eqnarray}
\!\!\!\!\!\!V &=& H_0^2\left[6J + \frac{2(3+32\xi)(1+12\xi)\xi}{(8\xi+1)^2}\sigma^2\right.\nonumber\\
&+& \left. \frac{4B_1B_0(1+12\xi)}{1+8\xi}\sigma^{-\frac{1}{4\xi}} -\frac{B_1^2B_0^2}{8\xi^2}\sigma^{-\frac{4\xi+1}{2\xi}}\right]\!.
\label{pot-old2}
\end{eqnarray}

Let us  apply the superpotential method to the same model with the de Sitter evolution.
Using~(\ref{SolF}),  we obtain
\begin{equation}
\label{F_deSitter}
F(\sigma) = \left[\int\limits^{\sigma} H_0e^\Upsilon d\tilde \sigma+ \tilde c_0\right] e^{-\Upsilon}
\end{equation}
and, in particular, for $U$, given by (\ref{particular}),
\begin{equation}
F(\sigma) = \frac{4\xi H_0}{8\xi+1}\sigma+\tilde c_0\sigma^{-\frac{1+4\xi}{4\xi}}.
\end{equation}
Therefore,
\begin{equation}
\sigma'=\tilde{F}(\sigma)=\frac{F(\sigma)}{Y(\sigma)}=\frac{4\xi}{8\xi+1}\sigma+\frac{\tilde c_0}{H_0}\sigma^{-\frac{1+4\xi}{4\xi}}.
\end{equation}
This formula coincides with (\ref{B1}), if we choose $\tilde c_{0}=H_{0}B_{1}B_{0}/(2\xi)$, so we conclude that the potential can be reconstructed with both the procedures.

Solving Eq. (\ref{equsigma}), we  obtain
\begin{equation}
\sigma(t)= \left[\sigma_0 e^{H_0 t}+\frac{\tilde c_0(8\xi+1)}{H_0\xi}\right]^{\frac{4\xi}{8\xi+1}},
\end{equation}
where $\sigma_0$ is an arbitrary constant.

For the  case $\tilde c_0=0$,  $F(\sigma)$ is a linear function.
The potential $V$ is the following quadratic polynomial:
\begin{equation}\label{pp}
V=2H_0^2\left[3J + \frac{(3+32\xi)(1+12\xi)\xi}{(8\xi+1)^2}\sigma^2\right].
\end{equation}
For $\xi=-1/12$ and $\xi=-3/32$ the potential
$V$ is a constant.

The case $\xi=-1/8$ should be considered separately. In this case, using Eq.~(\ref{SolF}), we get:
\begin{equation}
F(\sigma) = \sigma H_0\ln\pa{\frac{\sigma}{\sigma_{0}}},
\end{equation}
where $\sigma_{0}$ is an integration constant, and the corresponding potential has the following form:
\begin{eqnarray}
V&=&\frac{H_{0}^{2}}{4}\paq{24 J-\frac{\sigma^{2}}{2}\pa{\ln\pa{\frac{\sigma}{\sigma_{0}}}^{2}+3+\sqrt{3}}\right.\nonumber\\
&\times&\left.\pa{\ln\pa{\frac{\sigma}{\sigma_{0}}}^{2}+3-\sqrt{3}}}.
\end{eqnarray}
The scalar field evolution is given by
\begin{equation}
\sigma(t)=\sigma_{0}\exp\paq{{\displaystyle e^{H_0(t-t_0)}}}.
\end{equation}
It is easy to see that in the case of induced gravity ($J = 0$) the expressions (\ref{pot-old2}) and (\ref{pp}) can be derived from the expression (51) in the paper \cite{KTV11042125} provided we identify
$c_{1} = -\frac{B_1B_0(1+8\xi)}{8\xi^2}$ and $\gamma=2\xi$. Also the particular cases $\gamma=-1/6$ (conformal coupling) and $\gamma=-1/4$ corresponding to $\xi=-1/12$ and $\xi=-1/8$, respectively, were considered in \cite{KTV11042125}.

In the conclusion of this section, we would like to add that the exact solution with the potential (\ref{pot-old2})
was obtained independently by Starobinsky~\cite{Star-Mont0}. He has also shown that
the corresponding de Sitter solutions are stable only if the ``exotic terms'', depending on the fractional degrees
of the scalar field vanish due to the corresponding choice of the constants \cite{Star-Mont}.

\section{Power-law Solutions}
In general relativity with a barotropic perfect fluid, the Hubble parameter evolves as
\begin{equation}
H^2(a) = H^2_0\pa{\frac{a}{a_0}}^{-3(w+1)}=H_0^2 e^{nN}\,,
\label{bar1}
\end{equation}
where the constant $w$ is the equation-of-state parameter of the fluid and $n\equiv-3(w+1)$.
Let us note that the Hubble parameter as a function of time is $H(t)={}-\frac{2}{nt}=\frac{2}{3(w+1)t}$.
Power-law solutions for induced gravity models ($J=0$) and the corresponding potentials have been studied in~\cite{KTV11042125}. Let us note that, in the case with $J\neq 0$, the method employed in \cite{KTV11042125} leads to a differential equation that cannot be solved analytically.
For such a case it is convenient to adopt the superpotential technique for reconstruction.

Let us start from Eq. (\ref{solW}) and observe that the power-law dependence of $H^{2}$ on the scale factor gives $n=(H^{2})'/H^{2}$, or equivalently
\begin{equation} \label{ns}
n=\frac{W_{,\sigma} \tilde F}{W}.
\ee
Then Eq. (\ref{solW}) becomes
\begin{equation} \label{solWn}
\frac{n}{\tilde F}=\frac{\paq{2U_{,\sigma}-2\pa{U_{,\sigma\sigma}\tilde{F}+U_{,\sigma}\tilde{F}_{,\sigma}}-\tilde{F}}}
{\pa{2U+U_{,\sigma}\tilde{F}}}\,.
\ee

After substituting (\ref{particular}), Eq. (\ref{solWn}) can be recast in the following form:
\begin{eqnarray}
\label{solWn2}
&&4\xi\sigma \tilde F_{,\sigma}\tilde F+2\xi\sigma(n-2)\tilde F+(4\xi+1)\tilde F^2\nonumber\\
&&+2nJ+2n\xi\sigma^2=0.
\end{eqnarray}

In particular, for $n=2$, Eq.~(\ref{solWn2}) is a linear differential equation for $\tilde{F}^2$ that can be solved exactly leading to the following solution:
\begin{equation}
\label{F_n2}
  \tilde F(\sigma)=\pm \sqrt{c_1\sigma^{-\frac{4\xi+1}{2\xi}}-4\frac{(4\xi+1)\xi\sigma^2+(8\xi+1)J}{(4\xi+1)(8\xi+1)}}
\end{equation}
For this solution from Eq.~(\ref{ns}), we get
\begin{equation}
W(\sigma)=W_0e^{\int^{\sigma}\frac{n}{\tilde{F}(\tilde{\sigma})} d\tilde{\sigma}}.
\end{equation}

For the values of $\xi$ leading to a singular expression (\ref{F_n2}), Eq.~(\ref{solWn2}) has the following general solution:
\begin{itemize}
\item
for $\xi=-1/4$
\begin{equation}\label{Fxi1_4}
\tilde F(\sigma) = \pm\sqrt{8J\ln\left(\frac{\sigma}{\sigma_{0}}\right)-\sigma^2},
\end{equation}

\item
for $\xi=-1/8$
\begin{equation}\label{Fxi1_6}
\tilde F(\sigma) = \pm\sqrt{-2\sigma^2\ln\left(\frac{\sigma}{\sigma_{0}}\right)-8J}
\end{equation}
\end{itemize}
where $\sigma_{0}$ is the integration constant.

It is impossible to solve Eq.~(\ref{solWn2}) by the separation of $\sigma$ and $\tilde F$
for the case of an arbitrary $n$. Hence we look for particular solutions of Eq.~(\ref{solWn2}) by making a suitable ansatz for $\tilde F$ as a function of $\sigma$ and then algebraically fixing the free coefficients of the model.
We choose the ansatz
\begin{equation} \label{ansf1}
\tilde F(\sigma)=A_{0}+A_{1}\sigma\,,
\ee
by which Eq.~(\ref{solWn2}) reduces to a quadratic polynomial of $\sigma$, which can be solved by requiring that the coefficients of the diverse powers of $\sigma$ are zero.
Let us note that the above ansatz (\ref{ansf1}) when $A_{1}\neq 0$ corresponds to the following evolution for the scalar field in terms of $N$
\begin{equation} \label{sfevo1}
\sigma(N)=B\exp\pa{A_{1}N}-\frac{A_{0}}{A_{1}}
\ee
where $B$ is an integration constant. It is then straightforward to obtain the evolution $\sigma(t)$ by integrating:
\begin{equation} \label{NtoT}
dN=\frac{dN}{dt}dt=H dt=-\frac{2}{n\,t}dt\Rightarrow N=\ln\pa{\frac{t}{t_{0}}}^{-\frac{2}{n}}.
\ee
and substituting into Eq. (\ref{sfevo1}).
Equation~(\ref{solWn2}) gives three constraints on the five parameters of the model: $A_0$, $A_1$, $n$, $\xi$, and $J$. If we assume that the function $U(\sigma)$ is given by (\ref{particular}), or in other words that the parameters $\xi$ and $J$ are fixed, we then find the following expressions for the remaining three parameters:
\begin{equation}
\begin{split}
    A_0^2&=\frac{2\left[-14\xi-44\xi^2-1\pm(6\xi+1)\sqrt{52\xi^2+16\xi+1}\right] J}{(4\xi+1)^2\xi},\\
    A_1&=\frac{-36\xi^2-12\xi-1\pm(6\xi+1)\sqrt{52\xi^2+16\xi+1}}{(6\xi+1)(4\xi+1)},\\
   n&= \frac{14\xi+44\xi^2+1\mp(6\xi+1)\sqrt{52\xi^2+16\xi+1}}{(4\xi+1)\xi}.
\end{split}
\end{equation}

These solutions exist only if $\xi\neq -1/4$ and $\xi\neq-1/6$.
The parameters $n$ and $A_1$ are real if and only if $\xi<(-4-\sqrt{3})/26$ or $\xi>(-4+\sqrt{3})/26$.
Let us note that $A_0=0$ if and only if $J=0$.

On the other hand, if we assume that $n$ is given and seek both the functions $F$ and $U$, then Eq. (\ref{solWn2}) leads to two nontrivial solutions with $A_{0}$ unconstrained and
\begin{equation} \label{sol12m2}
{J}_{1,2}=A_{0}^{2}\frac{-4n-\alpha\pm\pa{n-2}\sqrt{\alpha}}{4n\pa{\alpha-2n}}
\ee
\begin{equation} \label{sol12xi}
\xi_{1,2}=\frac{8n-\alpha\mp\pa{n-2}\sqrt{\alpha}}{8\pa{\alpha-2n}}
\ee
\begin{equation} \label{sol12A1}
A_{1;1,2}=\frac{2-n \mp \sqrt{\alpha}}{4}
\ee
with $\alpha=4-20n+n^{2}$ and where $A_0$ and $-A_0$ correspond to the same values for the other parameters. These solutions describe a universe evolving according to (\ref{bar1}) due to the presence of a non-minimally coupled scalar field with the coupling to the Ricci scalar given by (\ref{sol12xi}). Note that $\alpha$ is non-negative and expressions (\ref{sol12m2})--(\ref{sol12A1}) are real provided $n\ge10+4\sqrt{6}\simeq 19.8$ or $n\le10-4\sqrt{6}\simeq 0.2$. Negative values of $n$ correspond to a decreasing $H$ in an expanding universe and describe relevant cosmological evolutions. The limit $n\rightarrow 0$ gives the de Sitter solution and the solutions with small, positive $n$ still describe interesting cosmological evolutions corresponding to a superaccelerated era (the fact that the Universe is currently undergoing a phase of superaccelerated expansion is not excluded by supernovae observations \cite{Gannouji:2006jm}).

Equation (\ref{solWn2}) also has four more solutions corresponding to either  the induced gravity or the general relativity framework.
We find induced gravity solutions (with $J=0$) having $A_{0}=0$, $\xi$ unconstrained and
\begin{equation} \label{sol34A1}
A_{1;3,4}=\frac{-(n-2)\xi\pm\sqrt{\xi\pa{\alpha \,\xi-2n}}}{1+8\xi}
\ee
or $\xi_{5}=-1/8$ and $A_{1;5}=n/\pa{2-n}$. In these cases the scalar field evolution is given by $\sigma(N)=B\exp{A_{1}N}$.
The general relativity solution is finally for  $A_{1}=0$, $\xi=0$, $J=-A_{0}^{2}/(2n)$ and a scalar field evolution given by $\sigma=A_{0}N+B$ where $B$ is an arbitrary constant. The constant $A_0$ is physically constrained by the value of
the Planck mass in the Einstein--Hilbert action.

Equation  (\ref{ns}) can be solved for $W$ with $\tilde F$ given by the ansatz (\ref{ansf1})
\begin{eqnarray}\label{solns0}
   W&=&W_{0}\exp\paq{\int^{\sigma}\frac{n}{\tilde F}d\tilde\sigma}
   =W_{0}\exp\paq{\int^{\sigma}\frac{n}{A_0+A_1\tilde\sigma}d\tilde\sigma} \nonumber\\
   &=&
   W_{0}\pa{A_0+A_1\sigma}^{\frac{n}{A_1}}.
\end{eqnarray}
where $W_{0}$ is an integration constant.

On using Eq. (\ref{recpN}), we obtain the corresponding potential
\begin{eqnarray} \label{recV}
V&=&W_{0}\paq{6\pa{J+\xi \sigma^{2}}+12\xi \sigma\pa{A_0+A_1\sigma}\phantom{\frac{B^2}{A}}\right.\nonumber\\
&-&\left.\frac{\pa{A_0+A_1\sigma}^{2}}{2}}
\pa{A_0+A_1\sigma}^{\frac{n}{A_1}}
\ea
with $J$, $\xi$ and $A_{1}$ given by (\ref{sol12m2})--(\ref{sol12A1}) respectively.
In the induced gravity case (solutions 3 and 4) the potential takes the following form:
\begin{equation} \label{recVpureIG}
V_{3,4}=\tilde W_{0}\sigma^{\frac{n}{A_{1;3,4}}+2}=\tilde W_{0}\sigma^{\frac{\pa{6-n}\xi\pm\sqrt{\xi\pa{\alpha\xi-2n}}}{2\xi}}
\ee
where we absorbed all the parameters and the integration constant $W_{0}$ into $\tilde W_{0}$. Correspondingly the evolution of the scalar field is given by
\begin{equation} \label{evscpureIG}
\sigma_{3,4}(N)=B\exp\paq{{\frac{-(n-2)\xi\pm\sqrt{\xi\pa{\alpha \,\xi-2n}}}{1+8\xi}N}}.
\ee
These two solution were already found in \cite{KTV11042125}.
In the $n\rightarrow 0$ (and $\alpha\rightarrow 4$) limit one then obtains the de Sitter solutions (see \cite{KTV11042125}) with
\begin{equation} \label{recVpureIGdS}
V_{3}^{(dS)}=\tilde W_{0}\sigma^{2}\;,\quad V_{4}^{(dS)}=\tilde W_{0}\sigma^{4}
\ee
and
\begin{equation} \label{evscpureIGdS}
\sigma_{3}^{(dS)}(N)=B\exp\paq{{\frac{4\xi}{1+8\xi}N}}\;,\quad \sigma_{4}^{(dS)}(N)=B.
\ee
If we  consider the induced gravity solution left, with $\xi_{5}=-1/8$, we are led to the potential
\begin{equation} \label{recVpureIG2}
V_{5}=\tilde W_{0}\sigma^{4-n}
\ee
and the evolution of the scalar field is given by
\begin{equation} \label{evscpureIG2}
\sigma_{5}(N)=B\exp\paq{{\frac{n}{2-n}N}}.
\ee
Let us note that such a solution cannot be obtained from the solutions 1--4 with an appropriate limit procedure.

The general relativity solution has
\begin{equation}\label{Wgr}
W=W_{0}\exp\pa{\frac{n}{A_{0}}\sigma}
\end{equation}
and a corresponding potential of the form
\begin{equation}\label{vgr}
V=\tilde W_{0}\exp\pa{\frac{n}{A_{0}}\sigma}
\end{equation}
which is the well-known exponential potential associated with power-law expansion.

The solutions 1 and 2 deserve some more discussion. In the limit $J_{1,2}\propto A_{0}\rightarrow 0$ one obtains induced gravity with the non-minimal coupling fixed at $\xi=\xi_{{1,2}}$. In such a case the potential takes the form
\begin{equation} \label{recVpureIG3}
V_{1,2}=\tilde W_{0}\sigma^{\frac{10-n\pm\sqrt{\alpha}}{4}}
\ee
and the scalar field evolves as
\begin{equation} \label{evscpureIG3}
\sigma_{1,2}(N)=B\exp\paq{\frac{2-n \mp \sqrt{\alpha}}{4}}.
\ee
Note that the solution 1, in the induced gravity limit, is the solution 4 with $\xi=\xi_{1}$ and the solution 2, in the same limit, is the solution 3 with $\xi=\xi_{2}$. Thus, when $J=0$, the solutions 1 and 2 are particular cases of the solutions~3 and 4.

\section{Solutions with the hyperbolic tangent}
Let us construct cosmological models, when the Hubble parameter is a function of the hyperbolic tangent. Such solutions are popular
in cosmology, because they can describe a bounce~\cite{Bounce} and late-time acceleration~\cite{AKV1,AKV2} having de Sitter/anti-de Sitter attractors both in the past and in the future. Such models can be reconstructed, for example, starting with the following ansatz for the scalar field evolution:
\begin{equation}
\label{tanh}
\sigma(t)=A\tanh\left[\omega(t-t_0)\right],
\end{equation}
where $A$, $\omega$, and $t_0$ are constants. Note that $t_0$ can be complex, so the parametrization (\ref{tanh}) includes the functions $\sigma(t)=A\coth\left[\omega(t-t_0)\right]$ as well.
For such functions Eq.~(\ref{equsigma}) takes the form
\begin{equation}
\label{Ftanh}
F(\sigma)=\omega\left(A-\frac1A\sigma^2\right).
\end{equation}
If $U(\sigma)=\xi\sigma^2$, then after substituting $F(\sigma)$, given by (\ref{Ftanh}), into formula (\ref{equasolY}) we can integrate the resulting equation to obtain
\begin{equation}
\label{Y_tanh_U2}
Y(\sigma)=\frac{(4\xi+1)\omega A}{8\xi\sigma}+\frac{(12\xi+1)\omega}{4\xi A}\sigma\ln\left(\frac{\sigma}{\sigma_{0}}\right)
\,,
\end{equation}
where $\sigma_{0}$ is an arbitrary constant.
The function $Y(\sigma)$ has the simplest form for $\xi=-1/12$:
\begin{equation}
\label{Y_C2_10}
Y(\sigma)=c_0\sigma-\frac{\omega A}{\sigma}.
\end{equation}
The corresponding potential is the fourth degree monomial:
\begin{equation}
V=-\frac{\left(\omega-A c_{0}\right)^{2}}{2 A^{2}}\sigma^4 \,.
\end{equation}
For $c_0=0$, we obtain the following evolution for the Hubble parameter:
\begin{equation} \label{htanh}
H={}-\omega\coth[\omega(t-t_0)]
\ee
and conversely we get $H=-\omega\tanh[\omega(t-t_0)]$, when $\sigma(t)=A\coth[\omega(t-t_0)]$.

In the more general case with $U(\sigma)=-\sigma^2/12+J$ we get
\begin{equation}
Y = c_0\sqrt{12J-\sigma^2}-\frac{\omega A}{6J}\sigma.
\end{equation}
For $c_0=0$, the Hubble parameter is proportional to $\sigma$ and the potential has the following polynomial structure:
\begin{eqnarray}
V&=&{}-\frac{(6J+A^2)^2\omega^2}{72J^2A^2}\sigma^4\nonumber\\
&+&\left(\frac{A^2}{3J}+1\right)
\omega^2\sigma^2
-\frac{1}{2}\omega^2 A^2.
\end{eqnarray}

Another way to get the desired Hubble parameter evolution (with constant $H$ attractors in the past and in the future) is to express $H$ as a function of $\sigma$ evolving as (\ref{tanh}) and then solve Eq.~(\ref{equa}) as a differential equation for $U(\sigma)$.
On assuming
\begin{equation}
\label{Ytanh1}
H=Y(\sigma)=B-C\sigma,
\end{equation}
where $B$ and $C$ are constants, and then substituting (\ref{Ftanh}) and (\ref{Ytanh1}) into Eq.~(\ref{equa}), we get the following second order linear differential equation for $U(\sigma)$:
\begin{eqnarray}
&&2\Omega (A^2-\sigma^2)U_{,\sigma\sigma}+2\paq{(C-2\Omega)\sigma-B}U_{,\sigma}\nonumber\\&&
-4CU+(A^2-\sigma^2)\Omega =0,
\end{eqnarray}
where we set  $\omega=\Omega A$.

A particular solution of this equation is the second degree polynomial
\begin{eqnarray}
\label{U_H_tanh}
U(\sigma)&=&{}-\frac{1}{12}\sigma^2+\frac{B}{6(2\Omega+C)}\sigma\nonumber\\
&+&\frac{2A^2C\Omega+4A^2\Omega^2-B^2}{12(2\Omega+C)C},
\end{eqnarray}
having the form (\ref{U2poly}).
For any values of $A$, $B$, $C$ and $\Omega$, except $\Omega=-C$, the potential $V(\sigma)$ is a fourth degree polynomial with a  coefficient of the high-order degree monomial equal to $-(\Omega+C)^2/2$.

Let us consider, for example, the particular set of parameters $B=A$ and $C=1$. We get
\begin{equation}
\label{U_H_tanh11}
U(\sigma)={}-\frac{1}{12}\sigma^2+\frac{A}{6(2\Omega+1)}\sigma+\frac{(2\Omega+4\Omega^2-1)A^2}{12(2\Omega+1)},
\end{equation}
hence, $U(\sigma)=0$ at
\begin{equation*}
\sigma_{1,2}=A\frac{1\pm 2\sqrt{2\Omega^2+2\Omega^3}}{2\Omega+1}.
\end{equation*}
It is easy to see that $U(A)>0$ for all $\Omega>0$, so, if we choose as a solution $\sigma(t)=A\tanh(\omega t)$, then
$U(t)$ is positive at late times if $A>0$.
We further observe that (see Fig. \ref{FigUV1}),
when $\Omega=1$, $U(\sigma(t))\geqslant0$ at any time because $-A\leqslant\sigma(t)\leqslant A$.
\begin{figure}[h]
\centering
\includegraphics[width=41mm]{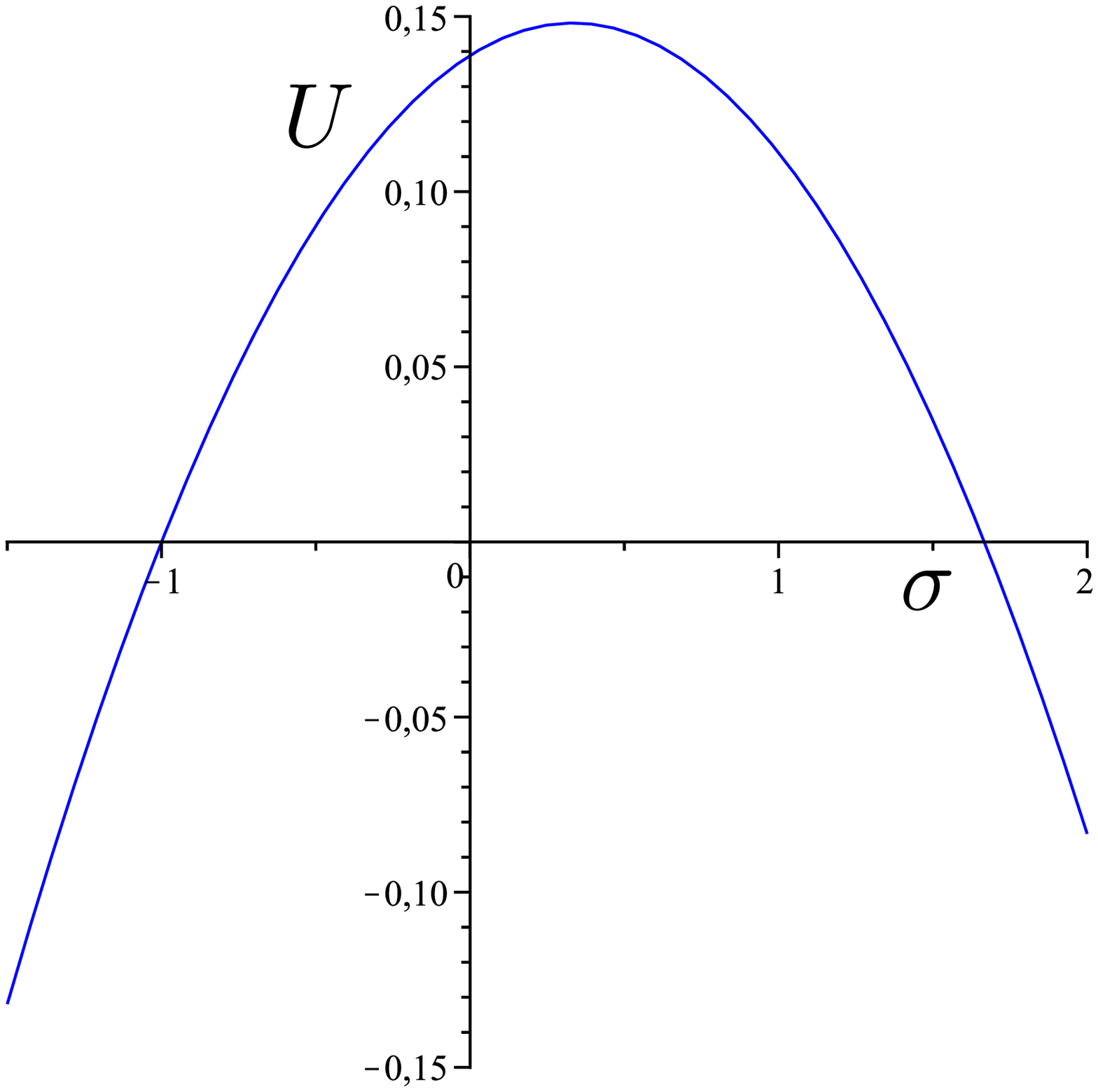}  \ \
\includegraphics[width=41mm]{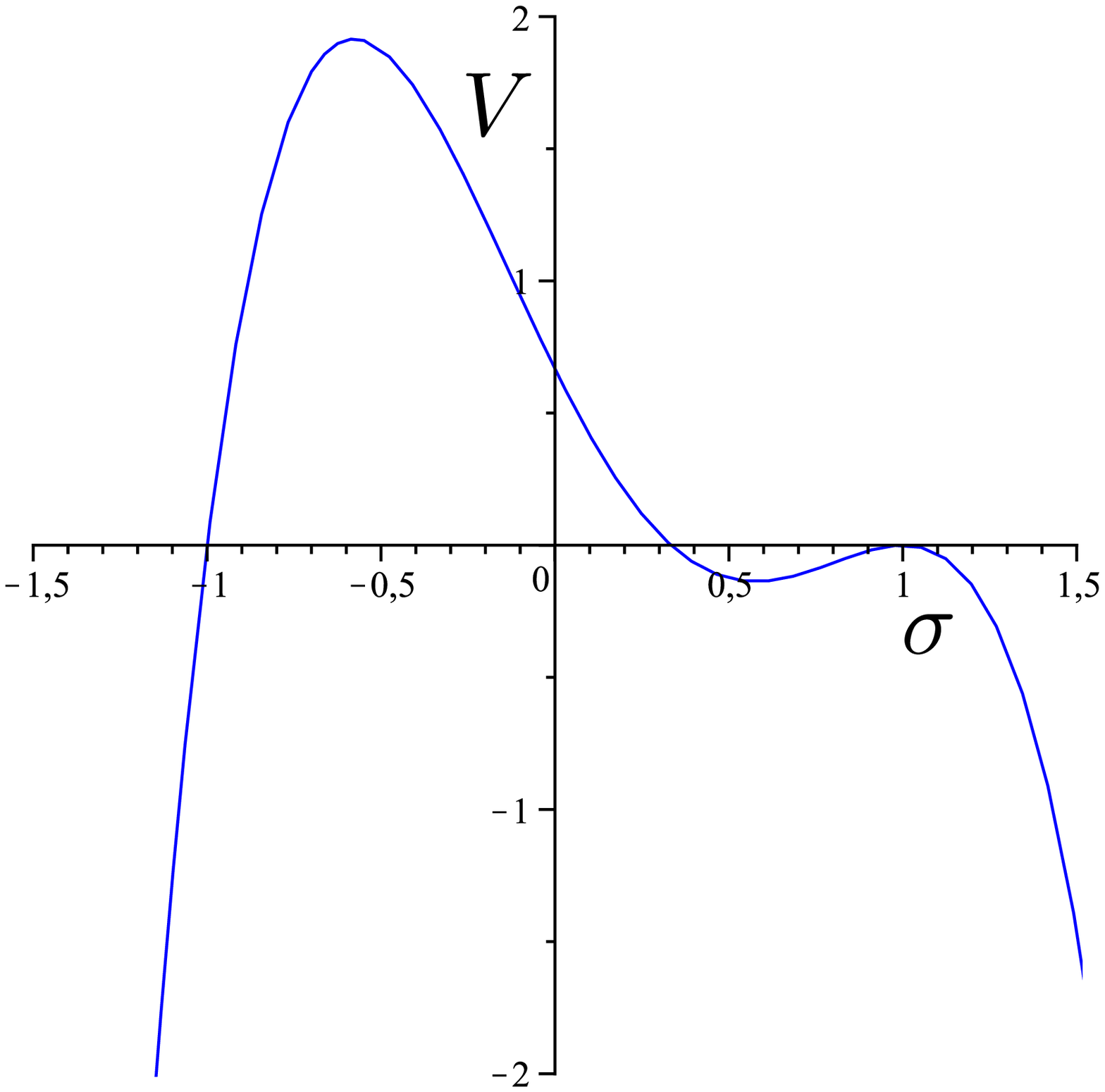}
\caption{The function $U(\sigma)$, given by (\ref{U_H_tanh11}), (left) and the potential $V(\sigma)$, given by (\ref{V_H_tanh}), (right) at $A=1$, $B=1$, $C=1$ and $\Omega=1$.} \label{FigUV1}
\end{figure}
For $\Omega=1$, we get
\begin{eqnarray}
\label{V_H_tanh}
V(\sigma)&=&{}-2\sigma^4+\frac{8}{3}A\sigma^3+\frac{4}{3}A^2\sigma^2-\frac{8}{3}A^3\sigma+ \frac{2}{3}A^4\nonumber\\
&=&{}-\frac{2}{3}(\sigma+A)(3\sigma-A)(\sigma-A)^2
\end{eqnarray}
and, by the change of variable $\tilde{\sigma}=\sigma-A/3$, the expression (\ref{U_H_tanh11}) becomes
\begin{equation}
U(\tilde\sigma)={}-\frac{1}{12}\tilde{\sigma}^2+\frac{4}{27}A^2,
\end{equation}
and takes the form of (\ref{particular}) with $\xi=-1/12$.
In terms of $\tilde{\sigma}$, we finally get
\begin{equation}
Y(\tilde\sigma)=\frac{2}{3}A-\tilde{\sigma},\quad F(\tilde\sigma)=\frac{\pa{2A-3\tilde \sigma}\pa{4A+3\tilde \sigma}}{9}.
\end{equation}

So, we found a model with exact solutions and $U(\sigma)$ in the form $U(\sigma)=\xi\sigma^2+J$.
We note that if we consider general values of $A$, $B$, $C$ and $\Omega$ in (\ref{U_H_tanh}) we still find a particular solution with $\xi=-1/12$. This value of $\xi$ corresponds to the case of the conformal coupling~\cite{KTV11042125}.

\section{Non-monotonic behavior of the Hubble Parameter in the Case of Induced Gravity}

In the previous sections we considered the case of $U(\sigma)$, specified by (\ref{particular}).
The  case of induced gravity ($J=0$) has been considered in detail in~\cite{KTV11042125}, where another method of reconstruction has been used. In this section, we find new induced gravity cosmological models with exact solutions on using the superpotential method. We demonstrate that the same evolution $\sigma(t)$ leads to exactly solvable models with different potentials and different qualitative behavior of the Hubble parameter.

Given $U(\sigma)=\xi\sigma^2$, and taking $Y(\sigma)$ as a generic polynomial
\begin{equation}
\label{Y2}
Y(\sigma)=\sum_{k=0}^N A_k\sigma^k,
\end{equation}
where $A_k$ are constants, from (\ref{equasolF}) one can obtain $F(\sigma)$ and show that it does not depend on $A_1$. For $N=2$, we obtain
\begin{eqnarray}
F(\sigma)&=&\frac{4\xi\left((16\xi+1)A_0-(8\xi+1)A_2\sigma^2\right)\sigma}{(8\xi+1)(16\xi+1)}\nonumber\\
&&+\tilde c_0\sigma^{-\frac{1+4\xi}{4\xi}}.
\end{eqnarray}
When $\tilde c_0=0$, $F(\sigma)$ is a cubic polynomial and Eq.~(\ref{equsigma}) has the general solution that can be written in terms of elementary functions
\begin{equation}\label{stt}
\sigma(t)=\pm\frac{\sqrt{(16\xi+1)A_0}}{\sqrt{(16\xi+1)A_0c_2e^{-\omega t}+(8\xi+1)A_2}}\,,
\end{equation}
where $\omega=8\xi A_0/(8\xi+1)$, $c_2$ is an arbitrary integration constant and we assume that $\xi\neq -1/8$.

The corresponding potential, $V(\sigma)$, is the sixth degree polynomial that, for example, when $\xi=1$, has the following form:
\begin{eqnarray}
V(\sigma)&=&\frac{910}{289}A_2^2\sigma^6+\frac{156}{17}A_1A_2\sigma^5\nonumber\\
&+&\left(6A_1^2+\frac{2236}{153}A_0A_2\right)\sigma^4+\frac{52}{3}A_0A_1\sigma^3\nonumber\\
&+&\frac{910}{81}A_0^2\sigma^2\,.
\end{eqnarray}

Let us analyze the cosmological consequences of the solution obtained.
If $\omega>0$, then
\begin{equation}
\lim_{t\rightarrow\infty}\sigma(t)=\pm\frac{(16\xi+1)\sqrt{A_0}}{(8\xi+1)\sqrt{A_2}}\,.
\end{equation}
In the case $\omega<0$, the function $\sigma(t)$ tends to zero at late times. Hence, the Hubble parameter  tends to a constant at late times for any case\footnote{We choose $Y(\sigma)$ to be a quadratic polynomial and therefore $A_2\neq 0$.}.
In Fig.~\ref{FigHY2} we  consider $H(t)$ for different values of $A_1$, for $\sigma$ positive, $\xi=1$, $A_2=1$, and $A_0=2$.
\begin{figure}[h]
\centering
\includegraphics[width=41mm]{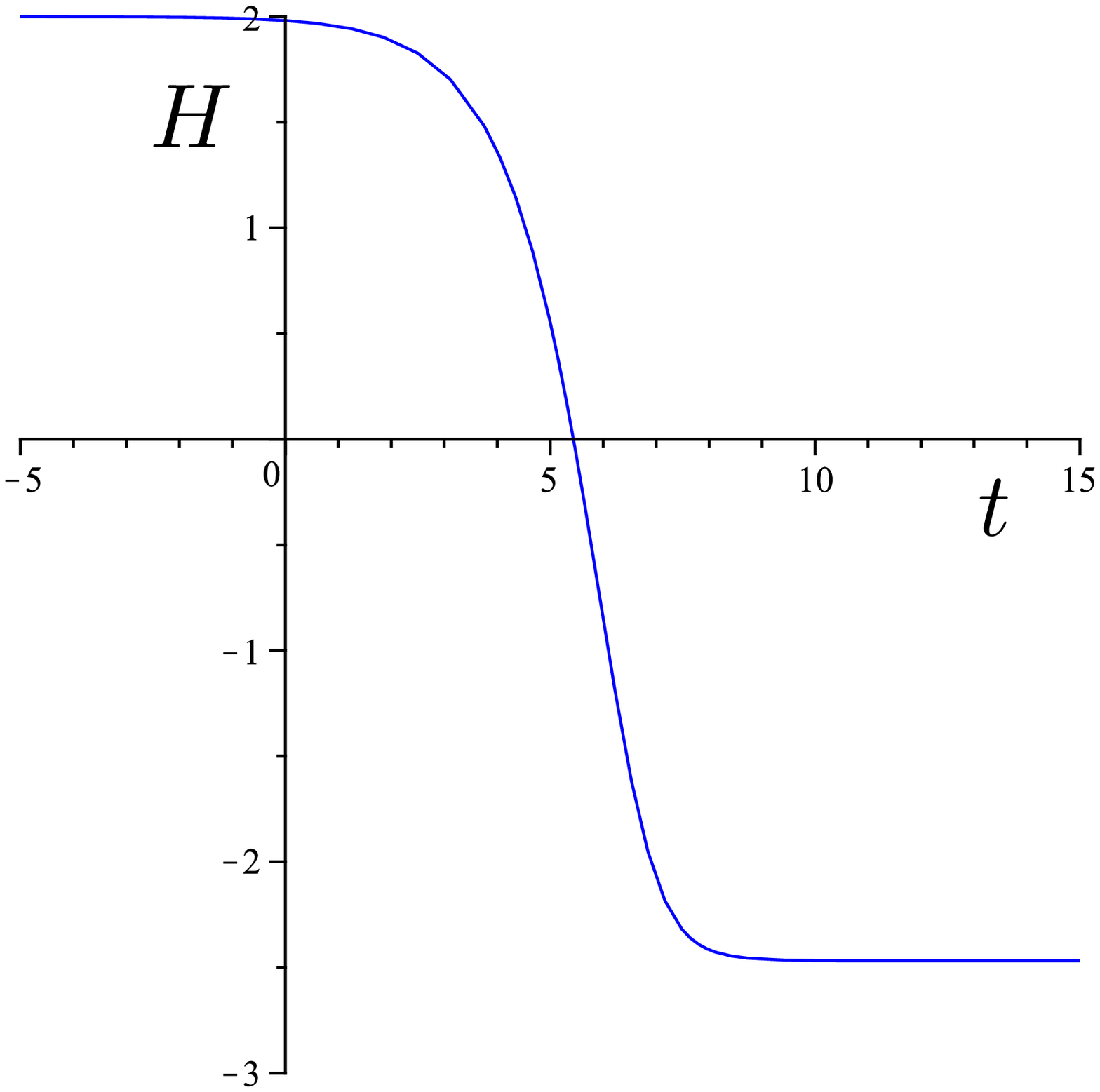} \ \
\includegraphics[width=41mm]{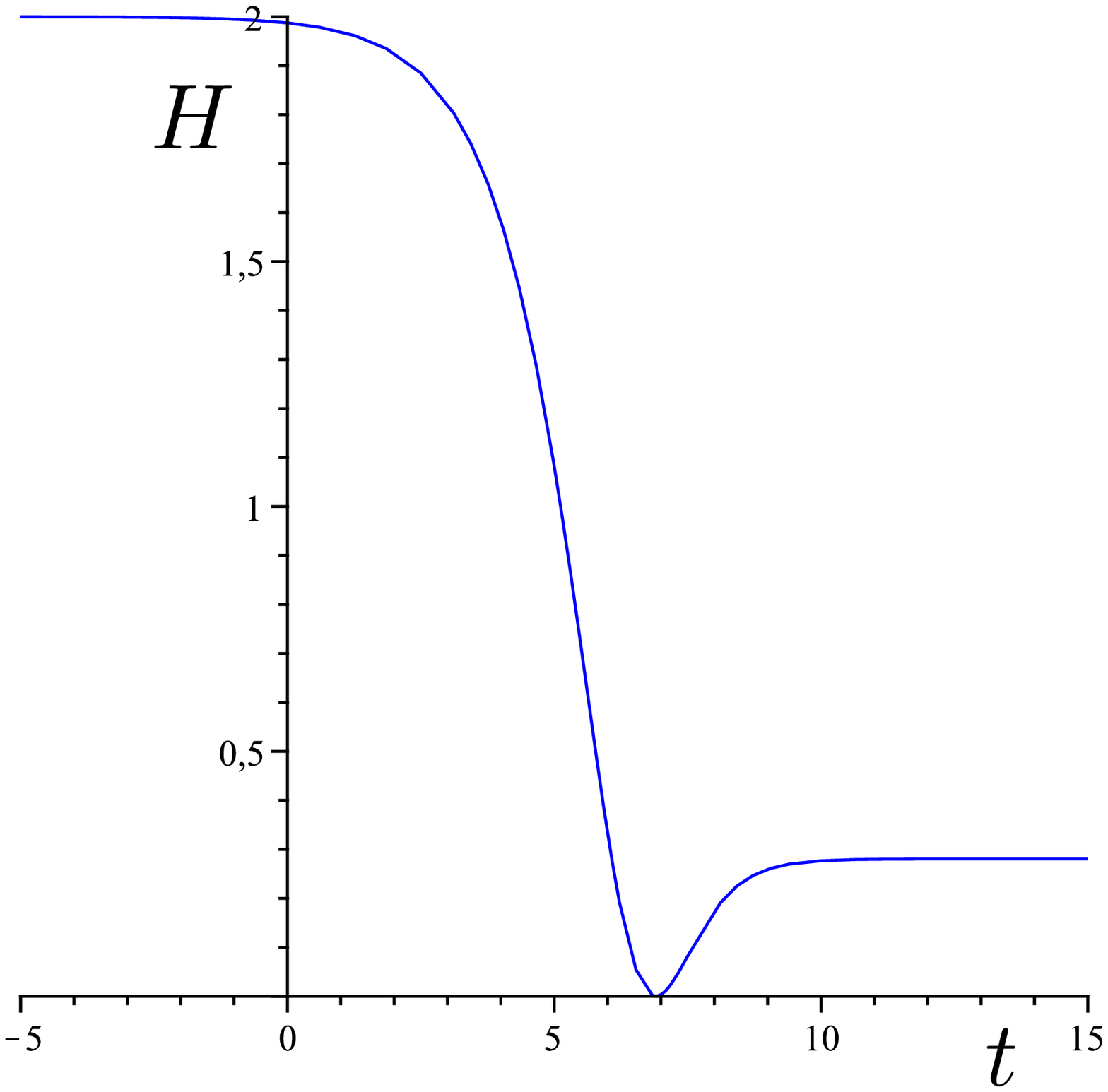} \ \
{ \ }\\
\includegraphics[width=41mm]{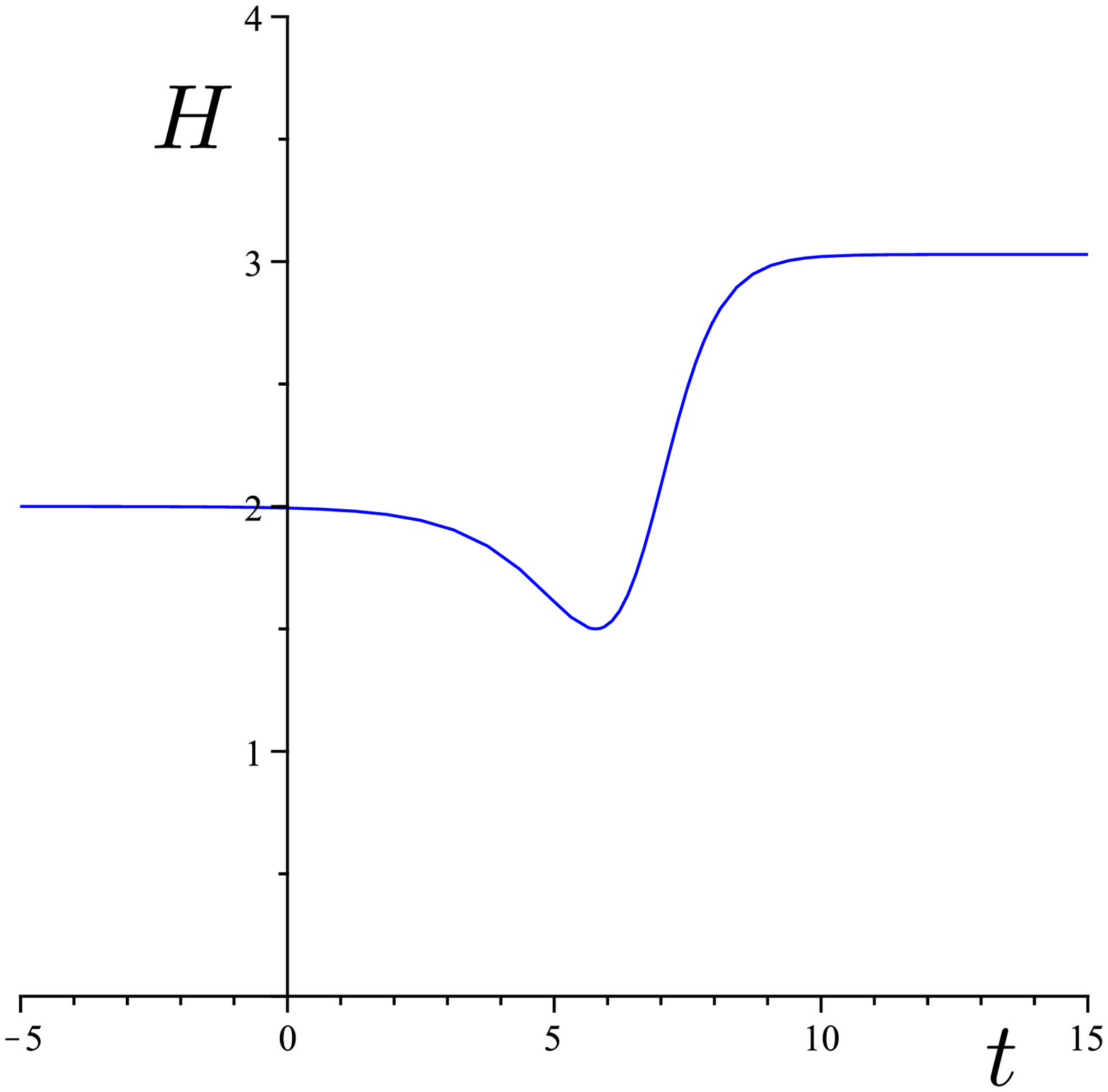} \ \
\includegraphics[width=41mm]{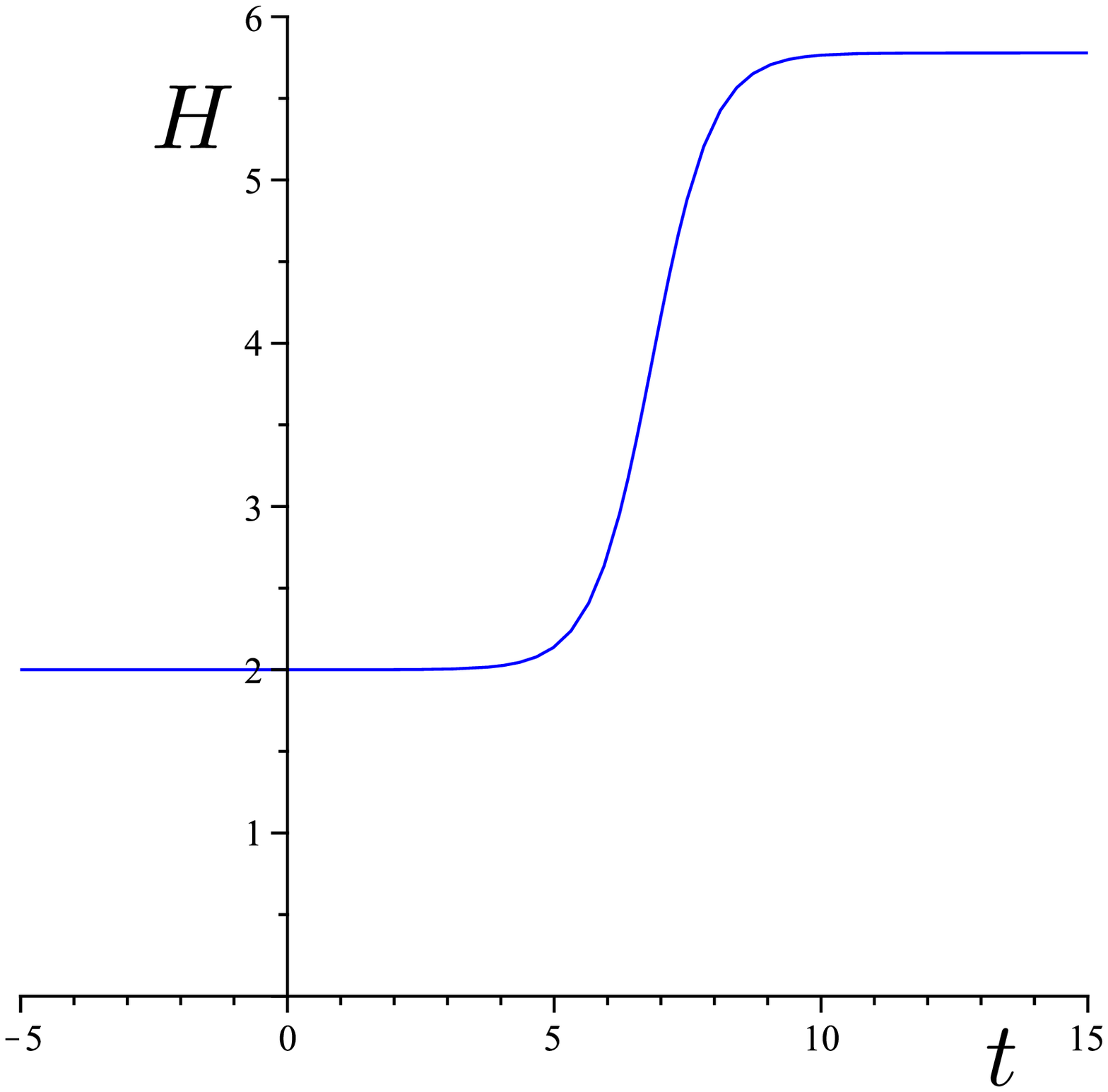}\\
\caption{The function $H(t)$, given by (\ref{Y2}), with $A_1=-6$ (upper plot on the left),
$A_1=-4$ (upper plot on the right), $A_1=-2$ (lower plot on the left), and $A_1=0$ (lower plot on the right).
At all pictures we use $A_2=1$, $A_0=2$,  and $c_2=100000$.} \label{FigHY2}
\end{figure}
The solution $\sigma(t)$ in (\ref{stt}) is associated with different behaviors of the Hubble parameter.
In particular, at $A_1=-4$ and $A_1=-2$ we get a non-monotonic behavior of $H(t)$. The behavior of the Hubble parameter, similar to the case $A_1=-4$, has been found in the quintom models with the sixth degree polynomial potential~\cite{AKV2}. Such behavior can, in principle, describe both inflation and the late-times accelerating expansion of the Universe and is thus physically relevant.

In this section, we have obtained new induced gravity models, with a polynomial potential and physically interesting behavior of the Hubble parameter.
Let us note that the explicit form of the Hubble parameter $H(t)$ is too complicated to be guessed and used in the reconstruction approach, proposed in~\cite{KTV11042125}.  We have obtained a three-parameter set of the induced gravity models with sixth degree polynomial potentials. The freedom of choice of parameters $A_i$ allows one to set additional conditions on the model.

\section{Construction of models with polynomial potentials}

On considering cosmological models with scalar fields, which are inspired by some fundamental theory, for example, string field theory, it is difficult to get the exact form of the potential, but it is possible to get, at least, some properties of the potential.
In~\cite{AKV1,AKV2} string field theory inspired models with minimally coupled scalar fields and polynomial potentials have been considered. Let us address the question regarding what functions $F$ and $U$ are associated with a polynomial potential.

To get a polynomial potential from formula (\ref{potentialV}) we assume that $F(\sigma)$ is an arbitrary $N$ degree polynomial:
\begin{equation}
F(\sigma)=\sum_{k=0}^N B_n\sigma^n.
\end{equation}

If $U=\xi\sigma^2$, then from Eq.~(\ref{equa}) we get
\begin{eqnarray}
Y(\sigma)&=&\frac {B_0(4\xi+1)}{8\xi\sigma}-\frac{B_2(12\xi+1)}{4\xi}\sigma\ln\left(\frac{\sigma}{\sigma_{0}}\right)\nonumber\\
&+&Y_P(\sigma),
\end{eqnarray}
where $Y_P(\sigma)$ is a polynomial. The term proportional to $\sigma^{-1}$ gives the constant term in the potential (\ref{potentialV}), thus only the term proportional to $\sigma\ln(\sigma)$ should be eliminated to obtain a polynomial potential.
Hence, we come to the conclusion that for an arbitrary polynomial $F(\sigma)$ we get a polynomial potential if $\xi=-1/12$.
This value of $\xi$ corresponds to the case of conformal coupling~\cite{KTV11042125}. For other values of $\xi$ we get a polynomial potential only if $B_2=0$.

If $U$ is an arbitrary quadratic polynomial, we get the following result: the coefficients of $Y(\sigma)$, proportional to $B_0$ and $B_1$,
are always polynomial, and those proportional to $B_2$ and $B_3$ are polynomial if $\xi=-1/12$. The coefficient
proportional to $B_4$ includes logarithmic terms. The function $Y(\sigma)$ also contains the term, proportional to $\sqrt{U(\sigma)}$, multiplied by an integration constant; therefore, to get a polynomial $V(\sigma)$ this integration constant should be zero. Hence, to get a polynomial potential, we should set either  $N=1$ or $N=3$ and $\xi=-1/12$. In the latter case,
we obtain a sixth degree potential for nonzero $B_3$. For $B_3=0$ (or equivalently $N=2$), the degree of the potential cannot be more than four.

\section{Conclusion}
In this article we employed the reconstruction procedure for cosmological models with non-minimally coupled scalar fields  evolving on a flat Friedmann-Robertson-Walker background. Besides their cosmological implications, models with non-minimally coupled scalar fields, including the Hilbert-Einstein term, are interesting because of their connection with particle physics. In particular we refer to models of inflation, where the role of the inflaton is played by the Higgs field non-minimally coupled to gravity. Such models have great relevance and were intensively discussed during the past years \cite{Higgs}.

It is difficult to apply the reconstruction procedure proposed in~\cite{KTV11042125} in the induced gravity context to more general modified gravity models.
For these models, with the superpotential method, one can find many exact solutions for the homogeneous scalar field-gravity system and the corresponding potential of the scalar field. Such a method translates the problem of integrating a second order differential equation with time dependent coefficients into the integration of a first order differential equation once either the Hubble parameter is expressed as a function of the scalar field or the behavior of the scalar field itself is given. Let us note that both the variants of the superpotential technique we are referring to do not need the expression of the Hubble parameter in terms of the cosmic time or of the scale factor. This method has been actively employed in cosmological models with minimally coupled scalar fields~\cite{Superpotential,AKV2,Andrianov:2007ua,ABV}, but, to the
best of our knowledge, it has not been applied to cosmological models with non-minimally scalar fields.
Hence the superpotential method is a powerful tool to find exact analytical solutions of cosmological interest.

In this article we used this method to find exact analytical solutions leading to some physically interesting behaviors of the Hubble parameter.
First we have found the potentials and the corresponding evolutions of the associated scalar field leading to de Sitter and power-law solutions. Power-law solutions reproduce the expansion of the Universe driven by a barotropic perfect fluid. For the former case we could formally reconstruct the dynamics of the scalar field-gravity system on assuming a generic coupling between the field and the Ricci scalar. In such a case both the superpotential technique and the reconstruction procedure proposed in~\cite{KTV11042125} can be applied. We then specialized the results to the non-minimally coupled case with a general coupling constant $\xi$.
The case of power-law expansion is more involved and we could only find and analyze some solutions with the parameters suitably tuned. Still, however, the reconstruction could only be performed easily by means of the superpotential technique.

We further investigated a few models having a de Sitter/anti de Sitter attractor far in the past or in the future.
In the non-minimal coupling case the dynamics of a scalar field with a suitable quartic potential has been shown to have these attractors.
In the induced gravity limit, we then also found a set of sixth degree polynomial potentials leading to a cosmic evolution with a constant $H(t)$ behavior both in the past and in the future.

Polynomial potential frequently appears as a result of the reconstruction procedure because polynomials are integrable and they can be systematically used for obtaining exact solutions with the superpotential technique. Hence we finally analyzed the sufficient conditions on the functions $U$ and $F$ to get a model with a polynomial potential.

To conclude,  let us compare our results with the results of other works, concerning the construction of exact solutions for models with a non-minimally coupled scalar field and sketch some possible future directions for investigation. In Ref.~\cite{CNP}  both the scalar potentials --- that of the self-interaction of the scalar field and that describing the interaction of the scalar field with the scalar curvature ---  were reconstructed. These potentials were chosen in such a way, that the cosmological evolution of the model reproduced the evolution of the $\Lambda$CDM model in the presence of radiation. The same class of potentials was also obtained on a purely theoretical ground, namely by imposing  maximal Noether symmetry on the scalar-tensor Lagrangian. Thus, the combination of the superpotential method
used in the present paper with the Noether symmetry method \cite{Rivista}  can represent a possible direction for the search of exact cosmological solutions.
Let us note that the simultaneous reconstruction of two scalar functions in a chameleon model, reproducing  evolution with the phantom divide line crossing was done in~\cite{cham}.

Another interesting question is a possible relation between the exact solutions for the models with the minimally coupled to gravity
scalar fields and  the models with the non-minimal coupling, which are connected by a conformal transformation. As is known, the question of the physical equivalence of the Jordan and Einstein frames is widely discussed in the literature (see e.g.
\cite{CL,Faraoni} and references therein), but at least mathematically, the transitions between these frames can be used as a useful
tool for the construction of new exact solutions.

Finally, we have tried to get some exact solutions with  background cosmological evolutions that are not far away from
that in the Standard Cosmological Model. The comparison of our solutions with the supernovae data can be done
immediately. As far as the comparison with the data on the microwave background cosmic radiation and the large
scale structure of the Universe  is concerned, the study of  cosmological perturbations is necessary, which is a rather complicated, but
challenging task. Some studies of the perturbations in modified gravity models were done in Ref.~\cite{Houri}.

\medskip

\noindent {\bf Acknowledgements.} We are grateful to A.A.~Starobinsky for useful discussions.
A.K. was partially supported by the RFBR grant 11-02-00643. The research of S.V. is supported in
part by the RFBR grant 11-01-00894, by the Russian Ministry of Education and Science under grant NSh-3920.2012.2,
and by contract CPAN10-PD12 (ICE, Barcelona, Spain).


\begin{thebibliography}{72}
\bibitem{cosmic}
A. Riess \textit{et al},
 \textit{Astron. J.}  \textbf{116} (1998) 1009--1038, arXiv:astro-ph/9805201; \\
S.J. Perlmutter  \textit{et al},
 \textit{Astroph. J.} \textbf{517} (1999) 565--586,  arXiv:astro-ph/9812133


\bibitem{DE_rev}
T. Padmanabhan, 
\textit{Phys. Rept.} \textbf{380} (2003) 235--320,
arXiv:hep-th/0212290;\\
P. Frampton,
\textit{Dark Energy -- a Pedagogic Review},
arXiv:astro-ph/0409166;\\
E.J. Copeland, M. Sami, and Sh. Tsujikawa,
\textit{Int. J. Mod. Phys. D} \textbf{15} (2006) 1753--1936, arXiv:hep-th/0603057;\\
 A.D. Dolgov,
\textit{Phys. Part. Nucl.} \textbf{43} (2012) 273--293;\\
 K. Bamba, S. Capozziello, S. Nojiri, and S.D. Odintsov,
 \textit{Astrophys. Space Sci.} 342 (2012) 155--228, arXiv:1205.3421;\\
V. Sahni and A.A. Starobinsky,
 \textit{Int. J. Mod. Phys.} D  \textbf{9} (2000) 373--444,  arXiv:astro-ph/9904398;\\
P.J.E. Peebles  and B. Ratra,
\textit{Rev. Mod. Phys.} \textbf{75} (2003) 559--606,  arXiv:astro-ph/0207347;\\
V. Sahni,
 \textit{Class. Quantum Grav.} \textbf{19} (2002) 3435--3448,  arXiv:astro-ph/0202076;\\
V.~Sahni and A.A.~Starobinsky,
\textit{Int. J. Mod. Phys.} D \textbf{15} (2006) 2105,  arXiv:astro-ph/0610026

\bibitem{Fujii_Maeda} Y. Fujii and K. Maeda,  \textit{The Scalar--Tensor Theory of Gravitation}, Cambridge University Press,
Cambridge, 2004

\bibitem{NO-rev}
 S.~Nojiri and S.D.~Odintsov,
\textit{Int.\ J.\ Geom.\ Meth.\ Mod.\ Phys.}  {\bf 4} (2007) 115--146, arXiv:hep-th/0601213;\\
S.~Nojiri and S.D.~Odintsov,
  \textit{Phys. Rept.} \textbf{505} (2011) 59--144, arXiv:1011.0544

\bibitem{Book-Capozziello-Faraoni}
S.~Capozziello and V.~Faraoni, \textit{Beyond Einstein Gravity: A Survey of Gravitational Theories for Cosmology and Astrophysics}, Fund. Theor. Phys. \textbf{170}, Springer, New York, 2011
\bibitem{CL}
S. Capozziello and M. De Laurentis, \textit{Phys. Rep.} {\bf 509} (2011) 167--321, arXiv:1108.6266

\bibitem{Felice_Tsujikawa}
A. de Felice and Sh. Tsujikawa,
\textit{Living Rev. Rel.} \textbf{13} (2010) 3, arXiv:1002.4928

\bibitem{inflation}
A.A.~Starobinsky,
\textit{Lect. Notes in Phys.} \textbf{246} (1986) 107;\\
 A.D.~Linde, \textit{Particle Physics and Inflationary Cosmology}, Chur, Switzerland: Harwood, 1990.
\bibitem{inflation2}
J.E. Lidsey, A.R. Liddle, E.W. Kolb, E.J. Copeland, T. Barreiro, and M. Abney,
\textit{Rev. Mod. Phys.} \textbf{69} (1997) 373--410,  arXiv:astro-ph/9508078;\\
C.M. Peterson, M. Tegmark,
\textit{Phys. Rev.} D \textbf{83} (2011) 023522, arXiv:1005.4056;\\
Shi Pi, M. Sasaki,
\textit{J. Cosmol. Astropart. Phys.} \textbf{1210} (2012) 051, arXiv:1205.0161


\bibitem{ENOSF} E. Elizalde, Sh. Nojiri, S.D. Odintsov, D. Sa\'ez-G\'omez, and V. Faraoni,
\textit{Phys. Rev.} D \textbf{77} (2008) 106005, arXiv:0803.1311




\bibitem{Starobinsky:1998fr}
  A.A.~Starobinsky,
  \textit{JETP Lett.}\   \textbf{68} (1998) 757--763,  arXiv:astro-ph/9810431

\bibitem{Burd:1988ss}
  A.B.~Burd and J.D.~Barrow,
  \textit{Nucl.\ Phys.}\  B  \textbf{308} (1988) 929--945
\bibitem{Barrow}
J.D.~Barrow,
  \textit{Phys.\ Lett.}\  B  \textbf{235} (1990) 40--43

\bibitem{Yurov:2003zt}
  A.~Yurov,
  \textit{Eur. Phys. J. Plus} \textbf{126} (2011) 132, arXiv:astro-ph/0305019
\bibitem{Guo:2006ab}
  Z.K.~Guo, N.~Ohta and Y.Z.~Zhang,
  \textit{Mod.\ Phys.\ Lett.}\  A  \textbf{22} (2007) 883,  arXiv:astro-ph/0603109
\bibitem{Guo:2005ata}
  Z.K.~Guo, N.~Ohta and Y.Z.~Zhang,
  \textit{Phys.\ Rev.}\  D \textbf{72} (2005) 023504,  arXiv:astro-ph/0505253



\bibitem{SRSS} T.D. Saini, S. Raychaudhury, V. Saini, and A.A.
Starobinsky,
\textit{Phys. Rev. Lett.} {\bf 85} (2000) 1162--1165, arXiv:astro-ph/9910231

\bibitem{AKV1}
I.Ya.~Aref'eva, A.S.~Koshelev, and S.Yu.~Vernov,
\textit{Theor. Math. Phys.} \textbf{148} (2006) 895--909, arXiv:astro-ph/0412619

\bibitem{scalarfields} S. Capozziello, S. Nojiri, and S.D. Odintsov,
\textit{Phys. Lett.} B \textbf{634} (2006) 93--100, arXiv:hep-th/0512118

\bibitem{Superpotential} A.G. Muslimov,
\textit{Class. Quant. Grav.} \textbf{7} (1990) 231--237,\\
D.S. Salopek and J.R. Bond,
\textit{Phys. Rev.} D \textbf{42} (1990) 3936--3962,\\
 V.M.~Zhuravlev, S.V.~Chervon and V.K.~Shchigolev,
  \textit{J.\ Exp.\ Theor.\ Phys.}   \textbf{87} (1998) 223;\\
D. Bazeia, C.B. Gomes, L. Losano, and R. Menezes,
\textit{Phys. Lett.} B \textbf{633} (2006) 415--419,
arXiv:astro-ph/0512197;\\
P.K. Townsend,
\textit{Class. Quant. Grav.} \textbf{25} (2008) 045017,
arXiv:0710.5178;\\
A.V. Yurov, V.A. Yurov, S.V. Chervon, and M. Sami,
\textit{Theor. Math. Phys.} \textbf{166} (2011) 259--269;\\
A.Yu. Kamenshchik and S. Manti,
\textit{Gen. Rel. Grav.} \textbf{44} (2012) 2205--2214, arXiv:1111.5183;\\
H.-Ch. Kim,
arxiv:1211.0604

\bibitem{NoOdin2007} S. Nojiri and S.D. Odintsov,
\textit{J. Phys. Conf. Ser.} \textbf{66} (2007) 012005, arXiv:hep-th/0611071


\bibitem{AKV2}
I.Ya. Aref'eva, A.S. Koshelev, and S.Yu. Vernov,
\textit{Phys. Rev.} D \textbf{72} (2005) 064017, arXiv:astro-ph/0507067;\\
S.Yu. Vernov,
 \textit{Theor. Math. Phys.} \textbf{155} (2008) 544--556, arXiv:astro-ph/0612487
\bibitem{Andrianov:2007ua}
  A.A.~Andrianov, F.~Cannata, A.Yu.~Kamenshchik, and D.~Regoli,
  \textit{J. Cosmol. Astropart. Phys.} {\bf 0802} (2008) 015, arXiv:0711.4300

\bibitem{ABV} I.Ya. Aref'eva, N.V. Bulatov, and S.Yu. Vernov,
\textit{Theor. Math. Phys.} \textbf{163} (2010) 788--803, arXiv:0911.5105

\bibitem{Padmanabhan:2002cp}
  T.~Padmanabhan,
  \textit{Phys.\ Rev.}\  D \textbf{66} (2002) 021301, arXiv:hep-th/0204150
\bibitem{Feinstein:2002aj}
  A.~Feinstein,
  \textit{Phys.\ Rev.}\  D  \textbf{66} (2002) 063511, hep-th/0204140

\bibitem{Gorini:2003wa}
  V.~Gorini, A.Y.~Kamenshchik, U.~Moschella, and V.~Pasquier,
  \textit{Phys.\ Rev.}\  D  \textbf{69} (2004) 123512,  arXiv:hep-th/0311111

\bibitem{Rotova}
V.K. Shchigolev and M.P. Rotova,
\textit{Mod. Phys. Lett.} A \textbf{27} (2012) 1250086, arXiv:1203.5030

\bibitem{Boisseau:2000pr}
  B.~Boisseau, G.~Esposito-Farese, D.~Polarski, and A.A.~Starobinsky,
  \textit{Phys.\ Rev.\ Lett.}\   \textbf{85} (2000) 2236,  arXiv:gr-qc/0001066


\bibitem{KTV11042125}
A.Yu. Kamenshchik, A. Tronconi, and G. Venturi,
  \textit{Phys. Lett.}  B \textbf{702} (2011) 191--196, arXiv:1104.2125

\bibitem{Qui}
T.~Qiu,
\textit{J. Cosmol. Astropart. Phys.} \textbf{1206} (2012) 041, arXiv:1204.0189;\\
T.~Qiu,
  \textit{Phys.\ Lett.\ B} {\bf 718} (2012) 475,
  arXiv:1208.4759


\bibitem{BNO}  K. Bamba, S. Nojiri, and S.D. Odintsov,
\textit{Phys. Rev.} D \textbf{77} (2008) 123532, arXiv:0803.3384;\\
E. Elizalde and A.J. L\'opez-Revelles,
\textit{Phys. Rev.} D \textbf{82} (2010) 063504, arXiv:1004.5021;\\
E. Elizalde, A.J. L\'opez-Revelles, S.D. Odintsov, and S.Yu. Vernov,
Cosmological models with Yang--Mills fields,
arXiv:1201.4302

\bibitem{delaCruzDombriz:2006fj}
A.~de la Cruz-Dombriz and A.~Dobado,
  \textit{Phys.\ Rev.\ } D {\bf 74} (2006) 087501,
  arXiv:gr-qc/0607118;\\
  A.~de la Cruz-Dombriz and D.~Saez-Gomez,
  \textit{Class. Quant. Grav.} 29 (2012) 245014, arXiv:1112.4481

\bibitem{FR_reconsruction}
Sh. Nojiri and S.D. Odintsov,
\textit{Phys. Rev.} D \textbf{74} (2006) 086005, arXiv:hep-th/0608008;\\
P.K.S. Dunsby, E. Elizalde, R. Goswami, S. Odintsov,  and D. Saez-Gomez,
\textit{Phys. Rev}. D  \textbf{82} (2010) 023519, arXiv:1005.2205;\\
R. Myrzakulov, D. Saez-Gomez, and A. Tureanu,
\textit{Gen. Rel. Grav.} \textbf{43} (2011) 1671--1684, arXiv:1009.0902


\bibitem{FT_Reconstruction}
M. Jamil, D. Momeni, M. Raza, and R. Myrzakulov,
 \textit{Eur. Phys. J.} C \textbf{72} (2012) 1999, arXiv:1107.5807;\\
K. Bamba, R. Myrzakulov, S. Nojiri, and S.D. Odintsov,
 \textit{Phys. Rev.} D \textbf{85} (2012) 104036, arXiv:1202.4057

\bibitem{DW} C.~Deffayet and R.P.~Woodard,
\textit{J. Cosmol. Astropart. Phys.} \textbf{0908} (2009) 023, arXiv:0904.0961

\bibitem{LocalScalarTensor}
T.S.~Koivisto,
  \textit{Phys.\ Rev.}  D {\bf 77} (2008) 123513,
  arXiv:0803.3399;\\
E. Elizalde, E.O. Pozdeeva, and S.Yu. Vernov,
\textit{Phys. Rev.} D \textbf{85} (2012) 044002, arXiv:1110.5806;\\
S.Yu. Vernov,
 \textit{Phys. Part. Nucl.} \textbf{43} (2012) 694--696, arXiv:1202.1172;\\
E. Elizalde, E.O. Pozdeeva, and S.Yu. Vernov,
\textit{Class. Quantum Grav.} \textbf{30} (2013) 035002, arXiv:1209.5957

\bibitem{Cooper:1982du}
  F.~Cooper and G.~Venturi,
  \textit{Phys.\ Rev.}\  D \textbf{24} (1981) 3338
\bibitem{induced}
 F. Finelli, A. Tronconi and G. Venturi,
  \textit{Phys. Lett.}  B \textbf{659} (2008) 466--470, arXiv:0710.2741
\bibitem{Cerioni:2009kn}
  A.~Cerioni, F.~Finelli, A.~Tronconi and G.~Venturi,
  \textit{Phys.\ Lett.}\  B  \textbf{681} (2009)  383--386, arXiv:0906.1902
\bibitem{Cerioni:2010ke}
  A.~Cerioni, F.~Finelli, A.~Tronconi and G.~Venturi,
  \textit{Phys.\ Rev.}\  D  \textbf{81} (2010) 123505, arXiv:1005.0935
 \bibitem{Tronconi:2010pq}
  A.~Tronconi and G.~Venturi,
  \textit{Phys.\ Rev.} D {\bf 84} (2011) 063517, arXiv:1011.3958

 \bibitem{Kamenshchik:2012rs}
  A.Y.~Kamenshchik, A.~Tronconi and G.~Venturi,
  \textit{Phys.\ Lett.} B {\bf 713} (2012) 358,
  arXiv:1204.2625

\bibitem{Sami:2012uh}
  M.~Sami, M.~Shahalam, M.~Skugoreva, and A.~Toporensky,
 \textit{Phys. Rev.} D \textbf{86} (2012) 103532, arXiv:1207.6691

\bibitem{CervantesCota:2010cb}
  J.L.~Cervantes-Cota, R.~de Putter, and E.V.~Linder,
  \textit{J. Cosmol. Astropart. Phys.} {\bf 1012} (2010) 019, arXiv:1010.2237

\bibitem{nonmin-inf}
B.~L.~Spokoiny,
 \textit{Phys.\ Lett.} B {\bf 147} (1984) 39;\\
T.~Futamase and K.-i.~Maeda,
 \textit{Phys.\ Rev.} D {\bf 39} (1989) 399;\\
D.S.~Salopek, J.R.~Bond and J.M.~Bardeen,
 \textit{Phys.\ Rev.} D {\bf 40} (1989) 1753;\\
R.~Fakir and W.G.~Unruh,
 \textit{Phys.\ Rev.} D {\bf 41} (1990)  1783



\bibitem{nonmin-quant}
A.O.~Barvinsky and A.Y.~Kamenshchik,
 \textit{Phys.\ Lett.}\ B {\bf 332} (1994) 270--276, arXiv:gr-qc/9404062;\\
A.O.~Barvinsky and A.Y.~Kamenshchik,
 \textit{Nucl.\ Phys.}\ B {\bf 532} (1998) 339--360, hep-th/9803052;\\
A.O.~Barvinsky, A.Y.~Kamenshchik, C.~Kiefer, and C.F.~Steinwachs,
 \textit{Phys.\ Rev.}\ D {\bf 81} (2010) 043530, arXiv:0911.1408

\bibitem{Chap}
A.Yu. Kamenshchik, U. Moschella and V. Pasquier, \textit{Phys. Lett.} B {\bf 511} (2001) 265

\bibitem{mod-Chap}
H.B. Benaoum, \textit{Accelerated universe from modified Chaplygin gas and tachyonic fluid},
  hep-th/0205140
\bibitem{we-Chap1}
A.Yu. ~Kamenshchik, U.~Moschella and V.~Pasquier,
  \textit{Phys.\ Lett.}\ B {\bf 487}  (2000) 7


\bibitem{Extradim}
A. Brandhuber and K. Sfetsos,
\textit{J. High Energy Phys.} \textbf{9910}
(1999) 013, arXiv:hep-th/9908116;\\
O. DeWolfe, D.Z. Freedman, S.S. Gubser, and A. Karch,
 \textit{Phys. Rev.}  D {\bf 62} (2000) 046008; arXiv:hep-th/9909134;\\
A.S. Mikhailov, Yu.S. Mikhailov, M.N. Smolyakov, and I.P. Volobuev,
\textit{Class. Quantum Grav.} \textbf{24} (2007) 231--242,
arXiv:hep-th/0602143;\\
K. Skenderis and P.K. Townsend,
\textit{Phys. Rev.} D \textbf{74} (2006) 125008, arXiv:hep-th/0609056
\bibitem{Star-Mont0}
A.A. Starobinsky, Exact de Sitter solutions with a non-constant scalar field in scalar-tensor gravity,
talk given at the Cosmology Worshop Montpellier12  on October 11, 2012.
\bibitem{Star-Mont}
A.A. Starobinsky, in preparation.
\bibitem{Bounce}
 T. Biswas, A. Mazumdar, and W. Siegel,
\textit{J. Cosmol. Astropart. Phys.} \textbf{0603} (2006)  009, arXiv:hep-th/0508194; \\
I.Ya. Aref'eva, L.V. Joukovskaya, and S.Yu. Vernov,
 \textit{J. Phys. A: Math. Theor.} \textbf{41} (2008) 304003, arXiv:0711.1364; \\
 T.~Biswas, T.~Koivisto, and A.~Mazumdar,
  \textit{J. Cosmol. Astropart. Phys.} {\bf 1011} (2010) 008, arXiv:1005.0590; \\
T. Biswas, A.S. Koshelev, A. Mazumdar, and S.Yu. Vernov,
\textit{J. Cosmol. Astropart. Phys.} \textbf{1208} (2012) 024, arXiv:1206.6374

\bibitem{Gannouji:2006jm}
  R.~Gannouji, D.~Polarski, A.~Ranquet, and A.A.~Starobinsky,
  \textit{J. Cosmol. Astropart. Phys.} {\bf 0609} (2006) 016, astro-ph/0606287
\bibitem{Higgs}
F.L.~Bezrukov and M.~Shaposhnikov,
 \textit{Phys.\ Lett.}\ B {\bf 659} (2008)  703--706, arXiv:0710.3755;\\
A.O.~Barvinsky, A.Y.~Kamenshchik and A.~A.~Starobinsky,
 \textit{J. Cosmol. Astropart. Phys.} {\bf 0811} (2008) 021, arXiv:0809.2104;\\
F.L.~Bezrukov, A.~Magnin and M.~Shaposhnikov,
 \textit{Phys.\ Lett.}\ B {\bf 675} (2009) 88--92, arXiv:0812.4950;\\
A.~De Simone, M.P.~Hertzberg and F.~Wilczek,
 \textit{Phys.\ Lett.}\ B {\bf 678} (2009)  1--8, arXiv:0812.4946;\\
F.~Bezrukov and M.~Shaposhnikov,
 \textit{J. High Energy Phys.} {\bf 0907} (2009)  089, arXiv:0904.1537;\\
A.O.~Barvinsky, A.Y.~Kamenshchik, C.~Kiefer, A.A.~Starobinsky and C.~Steinwachs,
 \textit{J. Cosmol. Astropart. Phys. }{\bf 0912} (2009) 003, arXiv:0904.1698;\\
F.~Bezrukov, A.~Magnin, M.~Shaposhnikov and S.~Sibiryakov,
 \textit{J. High Energy Phys.} {\bf 1101} (2011) 016, arXiv:1008.5157;\\
A.O.~Barvinsky, A.Y.~Kamenshchik, C.~Kiefer, A.A.~Starobinsky and C.F. ~Steinwachs,
\textit{Eur. Phys. J.} C  {\bf 72} (2012) 2219, arXiv:0910.1041

\bibitem{CNP}
S. Capozziello, S. Nesseris and L. Perivolaropoulos, \textit{J. Cosmol. Astropart. Phys.} {\bf 0712} (2007) 009,
arXiv:0705.3586

\bibitem{Rivista}
S. Capozziello, R. de Ritis, C. Rubano and S. Scudellaro, \textit{Riv. Nuovo Cimento} {\bf 19}, N4 (1996) 1--114

\bibitem{cham}
F. Cannata and A.Yu. Kamenshchik, \textit{Int. J. Mod. Phys.} D {\bf 20} (2011) 121--131, arXiv:1005.1878

\bibitem{Faraoni} V. Faraoni and Sh. Nadeau,
\textit{Phys. Rev.} D \textbf{75} (2007) 023501,  arXiv:gr-qc/0612075
\bibitem{Houri}
H.~Ziaeepour,
  \textit{Phys.\ Rev.}\ D {\bf 86} (2012) 043503, arXiv:1112.6025
\end{thebibliography}
\end{document}